\documentclass[pre,nofootinbib,floatfix, superscriptaddress,a4paper]{revtex4}
\usepackage{amsmath,bm,graphicx}
\usepackage{amssymb}
\usepackage{amsfonts}
\usepackage[normalem]{ulem}
\DeclareMathOperator{\arctanh}{arctanh}
\usepackage[dvipsnames]{color}
\usepackage{natbib}
\newcommand{\pa}[1]{\textcolor{black}{#1}}

\begin{document}

\title{
\pa{Model} microswimmers in channels with varying cross section
}

\author{Paolo Malgaretti}
\email{malgaretti@is.mpg.de}
\affiliation{Max-Planck-Institut f\"{u}r Intelligente Systeme, Heisenbergstr. 3, D-70569 Stuttgart, Germany}
\affiliation{IV. Institut f\"ur Theoretische Physik, Universit\"{a}t Stuttgart, Pfaffenwaldring 57, D-70569 Stuttgart, Germany}

\author{Holger Stark}
\affiliation{Institut f\"ur Theoretische Physik, Technische Universit\"at Berlin, Hardenbergstrasse 36, 10623, Berlin, Germany}

\date{\today}

\begin{abstract}
We study different types of microswimmers moving in channels with varying cross section and thereby interacting
hydrodynamically with the channel walls. Starting from the Smoluchowski equation for a dilute suspension, for which interactions among swimmers can be neglected, we derive 
analytic expressions for the lateral probability distribution between plane channel walls. For weakly corrugated channels we extend 
the Fick--Jacobs approach to microswimmers and thereby derive an effective equation for the probability distribution along
the channel axis. Two regimes arise dominated either by entropic forces  due to the geometrical confinement 
or by the active motion.
In particular, our results show that the accumulation of microswimmers at channel walls is sensitive to both, the underlying swimming mechanism and the geometry of the channels. Finally, for asymmetric channel corrugation our model predicts a rectification of microswimmers along the channel, the strength and direction of which strongly depends on the swimmer type.
\end{abstract}

\maketitle

\section{Introduction}
Organisms as well as synthetic particles swimming at low Reynolds number attain net displacement by locally stirring the fluid~\cite{LaugaPowers}. 
Therefore, their swimming performances are affected by the presence of boundaries, interfaces, or other particles, which perturb the fluid flow generated by the microswimmers and thereby ultimately influence their swimming speed.
For example, sperm cells as well as bacteria, have been shown to accumulate at solid walls~\cite{Rotschild1963,DiLuzio2005,Lauga2006,Friedrich2010}. Moreover, in the presence of solid boundaries or fluid-fluid interfaces E.Coli bacteria swim along circular trajectories~\cite{Berg1990,DiLeonardo2011,chinappi2016}. 
In many biological situations as well as technological applications microswimmers are required to move in confined regions as it happens in  microfluidic devices or in the female reproductive tract~\cite{Suarez}.  
Recently, such situations have been addressed more systematically both in theory~\cite{Popescu2009,Spagnolie2012,ElgetiReview,Uspal2015,Colberg2015,Wu2015,Dominguez2016,Dominguez2016-2,Stark2016,Malgaretti2016,LoewenDDFT} as well as experiments~\cite{Hulme2008,Drescher2011,Denissenko2012,Clement}. 

Up to now, much attention has been payed to the case in which active particles are confined by walls or interfaces. However, for passive systems it is well known that the shape of the boundaries can induce novel 
dynamic regimes absent in the case of homogeneous confinement~\cite{HanggiReview,BuradaReview,Malgaretti2013}.
Similarly, recent works have shown that active particles moving in a non-uniform geometrical confinement 
give rise to novel dynamic effects.
For example, in experiments the motion of bacteria was rectified by funnel~\cite{Galajda2007} or ratchet-like~\cite{DiLeonardo2010} potentials. Moreover, theoretical models dealing with active particles under the influence of
period enthalpic~\cite{Pototsky2013} or entropic~\cite{Ghosh2013,Ghosh2014,Wu2015} potentials, show rectification 
of active motion under suitable conditions. 
Or, they demonstrate how active Brownian and run-and-tumble particles separate in circular mazes \cite{Khatami2016}.
Finally, dynamic collective effects 
arise, for example, when bacteria are confined in circular pools, where they spontaneously circle in one direction~\cite{Lushi2014}, or when active particles are confined in narrow channels joining larger reservoirs, where self-sustained density oscillations 
evolve~\cite{Paoluzzi2015}. 

\begin{figure}
 \includegraphics{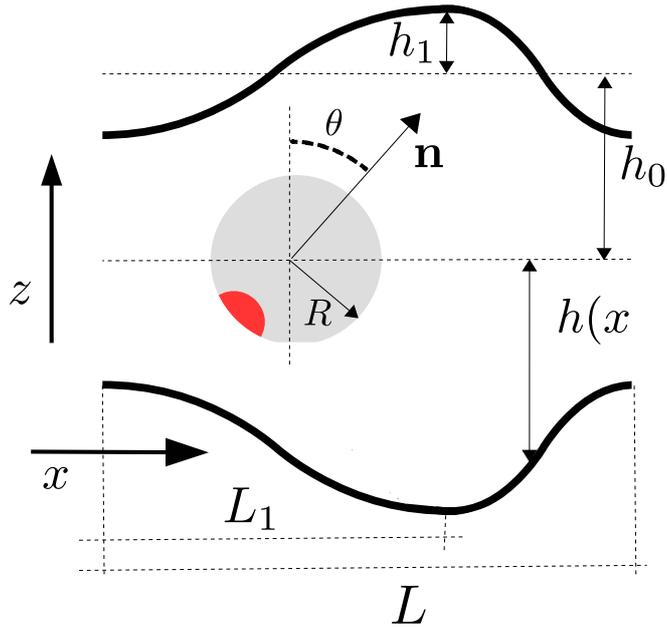}
 \caption{Schematic view of a microswimmer confined between corrugated walls. \pa{For the sake of clarity the amplitude of the modulation of the corrugated channel walls is strongly exaggerated.}}
\label{fig.1}
\end{figure}

In this contribution, we study the dynamics of microswimmers moving in a channel with varying cross section,
where channel walls are corrugated.
In particular, approximating the fluid flow, initiated by a microswimmer, by the leading far--field expression~\cite{Drescher2011,Spagnolie2012,Schaar2015}, we derive an effective interaction between microswimmers 
and the corrugated channel walls that depends on both the swimming mechanism and the wall geometry.
In order to derive analytical expressions, we assume that the channel cross section varies smoothly enough such that swimmers explore the cross-sectional direction while moving along the channel axis.
In such a regime, the Fick--Jacobs  approximation~\cite{zwanzig,Reguera2001,berezhkovskii2007diffusion,kalinay2008,Berezhkovskii2011,dagdug2013-2,dagdug2013} 
properly captures the overall dynamics of systems as diverse as molecular motors~\cite{Malgaretti2012,Malgaretti2013confined,MalgarettiEPJST}, driven particles~\cite{Reguera2006,Reguera2012}, electrokinetic phenomena~\cite{Malgaretti2014,Malgaretti_macromolecules,Malgaretti2015}, and polymer translocation~\cite{Bianco2016} just to mention a few among others (see recent review articles~\cite{HanggiReview,BuradaReview,Malgaretti2013} for a more comprehensive list). 

Here, we extend the Fick--Jacobs approximation to active particles\footnote{A first attempt to generalize the Fick--Jacobs equation to the case of active Brownian particles (i.e. in the absence of hydrodynamic coupling) can be found in Ref.~\cite{Dagdug2014}} and we will characterize the motion of microswimmers in narrow
channels with corrugated walls. In particular, our model shows that the geometrical confinement and the hydrodynamic interactions of the microswimmer with the channel walls can induce opposing effective forces on 
the swimmer. Hence, there is a cross-over value of the P\'eclet number, for which these two contributions are equal in magnitude. 
\pa{The whole formalism is valid independent of a special swimmer type.}
In order to quantify the effective coupling between the geometrical confinement and the microswimmers, we use 
generic expressions of the flow field induced by \pa{model} microswimmers
\pa{such as pushers, pullers, and source dipoles similar to the approaches in  Refs.~\cite{Ortiz2005,LaugaPRL2008,Baskaran15092009,Spagnolie2012,Lauga2014,Hennes2014,Schaar2015}}.
The no-slip boundary condition at the channel walls is accounted for by the method of images~\cite{Spagnolie2012}. 
Our results show that the strength of accumulation of the swimmers at the channel walls is strongly affected by both the channel 
geometry and the underlying swimming mechanism of pushers, pullers, and source dipoles. 
In particular, fast pullers accumulate at different locations along the channel than pushers and source dipoles, hence allowing for a mechanism to separate active particles. 
Finally, for asymmetric channels we show that the motion of the swimmers becomes rectified and that different swimmer types can 
move in opposite directions along the channel axis.

\pa{
The article is organized as follows. In section II we derive general expressions for our model of confined microswimmers. In sections II.A and II.B we then take the limit of small ratio of rotational to translational P\'eclet number, which means narrow channels, and apply our model to the two cases of plane walls and smoothly corrugated walls. In order to exploit the predictions of the model, in section III we use the generic expressions for the flow fields of model microswimmers to discuss their 
dynamics in channels with plane and corrugated walls. Finally, in section IV we summarize our results.}

\section{Model}

We analyze the dynamics of spherical swimmers moving between two corrugated walls, whose distance $2h(x)$ varies solely 
along the $x$ direction  and is constant along the $y$ direction (see Fig.\ \ref{fig.1}).
In order to keep our analysis simple and to be able to formulate analytical insight, we assume that the swimming axes of the active particles is constrained to the $x-z$ plane\footnote{Experimentally such a configuration can be attained by trapping the swimmers by means of optical tweezers.} and that the system 
extends infinitely along the $y$ direction. Such an assumption together with the homogeneity of the channel cross section along  the $y$ axis makes the probability distribution independent of $y$ and the problem becomes two-dimensional.
Accordingly, in the overdamped regime, the dynamics of a dilute suspension of non-interacting microswimmers is captured by the  Smoluchowski equation that governs the temporal evolution of the density $\rho(\mathbf{r},\theta)$ of swimmers located at $\mathbf{\mathbf{r}}=(x,z)$ and with their swimming  axes oriented at an angle $\theta$ against the $z$ axis~\cite{Elgeti2013}:
\begin{equation}
\frac{\partial 
}{\partial t} \rho (\mathbf{r},\theta)=-\vec{\nabla}\cdot\mathbf{J} (\mathbf{r},\theta)
\label{eq:smol_1}
\end{equation}
where 
\begin{equation}
\mathbf{J}=\left(\begin{array}{c}
\mathbf{J}_{\mathbf{r}}\\
J_{\theta}
\end{array}\right)
\qquad \mathrm{and} \qquad
\vec{\nabla} = \left( \begin{array}{c}
\vec{\nabla}_{\mathbf{r}}\\
\frac{\partial}{\partial \theta}
\end{array}\right)
\label{eq:smol_2}
\end{equation}
with
\begin{eqnarray}
\mathbf{J}_{\mathbf{r}}(\mathbf{r},\theta) & = & 
 [ -D_{t} \vec{\nabla}_{\mathbf{r}} + v_{0} \left(\mathbf{n}+\mathbf{v}(\mathbf{r},\theta)\right) - \beta D_{t}  \vec{\nabla}_{\mathbf{r}} C(\mathbf{r})] \rho(\mathbf{r},\theta) 
\label{eq:def-linear-flux}\\
J_{\theta}(\mathbf{r},\theta) & = &
\left[ -D_{r}\frac{\partial}{\partial \theta} + \omega_{0} \omega(\mathbf{r},\theta) \right] \rho(\mathbf{r},\theta) \, .  
\label{eq:def-ang-flux}
\end{eqnarray}
Here, $\mathbf{J}_{\mathbf{r}}$ and $J_\theta$ are the respective translational and rotational probability current
densities,
which consist of diffusional ($D_t$, $D_r$), active ($v_0 \mathbf{n}$), 
drift ($-\beta D_t \vec{\nabla}_{\mathbf{r}} C$), and wall-induced ($v_0 \mathbf{v}$, $\omega_0\omega$)
contributions.
In particular, $D_t$ ($D_r$) is the translational (rotational) diffusion coefficient\footnote{We remark that when a particle 
approaches
a wall, its 
translational and rotational
diffusion coefficients
become anisotropic and have to be described by tensors, the coefficients of which depend on the distance to
the bounding wall.
The
anisotropy is due to the 
different
hydrodynamic coupling perpendicular and parallel to the wall. 
Furthermore, one also has a
wall-induced 
rotation-translation
coupling
tensor.
In the following we are interested in the regime of moderate P\'eclet numbers
with $\mathrm{Pe}_t  > 1$,
for which diffusion is subdominant as compared to the active displacement of the microswimmers. Accordingly, in order to keep the derivation of the model as simple as possible, we neglect these corrections in the diffusion tensors and we regard $D_t$ and $D_r$ as constant scalar parameters.}, $v_0 \mathbf{n}$ is the swimming velocity directed along the unit vector $\mathbf{n}$, and $\beta^{-1}=k_BT$ with $k_B$ the Boltzmann constant and $T$ the absolute temperature. The dimensionless quantities $\omega$ and $\mathbf{v}$ are the respective angular and linear velocities induced by the hydrodynamic coupling
of the swimmer flow field with the channel walls (see appendix\ \ref{app:3} for explicit expressions)
and $\omega_0=v_0/R$ is the relevant angular-velocity scale with particle radius $R$. We remark that for
active Brownian particles, where any hydrodynamic interactions with the walls are neglected, one has $\omega=0$ and 
$\mathbf{v} = \mathbf{0}$, while for passive particles, for which $v_0=\omega_0=0$, Eqs.~(\ref{eq:def-linear-flux}) and 
(\ref{eq:def-ang-flux}) reduce to the standard diffusion equations
including a translational drift current induced by the geometrical confinement.

The force potential $C$ in Eq.~(\ref{eq:def-linear-flux}),
\begin{equation}
C(\mathbf{r})=C(x,z)=\begin{cases}
0 & |z|\leq h(x)-R+\mathcal{O}( | \partial_x h(x) |^2)\\
\infty & |z|>h(x)-R+\mathcal{O}( | \partial_x h(x)|^2) \, ,
\end{cases}
\label{eq:def_C}
\end{equation}
confines particles between the channel walls located at  $\pm h(x)$\footnote{Eq.~(\ref{eq:def_C}) has been obtained by expanding the exact expression for the available space a rigid sphere of radius $R$ can explore in a channel with varying cross section about a channel with constant cross section, $h(x)=h_0$, assuming small values of $\partial_x h(x)$.}.

The steady-state probability density is obtained by solving
\begin{equation}
\vec{\nabla}\cdot\mathbf{J}=\vec{\nabla}_{\mathbf{r}}\cdot\mathbf{J}_{\mathbf{r}}+
\frac{\partial}{\partial \theta } J_{\theta} = 0 \, ,
\label{eq:smol_steady_state}
\end{equation}
which gives
\begin{equation}
\nabla_{\mathbf{r}}^{2}\rho +   
\vec{\nabla}_{\mathbf{r}}\cdot [\rho\vec{\nabla}_{\mathbf{r}}\beta C ]
- \mathrm{Pe}_t \vec{\nabla}_{\mathbf{r}}\cdot\left[\left(\mathbf{n}+\mathbf{v}\right)\rho\right]
+ \frac{\mathrm{Pe}_t}{\mathrm{Pe}_r}
\frac{\partial}{\partial\theta}\left[\left(\frac{\partial}{\partial\theta}-\frac{\omega_0}{D_r}\omega\right)\rho \right]
=0 \, , 
\label{eq:smol_7}
\end{equation}
where we normalized all lengths by $d=h_0-R$. 
Furthermore, we identified the rotational, 
$\mathrm{Pe}_{r} =\frac{v_{0}}{D_{r}d}$, and translational, $\mathrm{Pe}_{t}=\frac{v_{0}d}{D_{t}}$, P\'eclet numbers. 
Equation~(\ref{eq:smol_7}) is complemented by the following
boundary conditions using the unit vector $\mathbf{n}_h$ for the normal at the channel walls:
\begin{eqnarray}\label{eq:div-J}
   \mathbf{n}_h(x) \cdot  \left. \mathbf{J}_{\mathbf{r}} \right|_{\pm \frac{h(x)-R}{d}} =0 
   && \text{\ldots \,no flux across channel walls}\label{eq:BC-Jz}\\
 \rho(\mathbf{r},\theta)=\rho(\mathbf{r}+L\mathbf{e}_x,\theta) && \text{\ldots \,periodicity along channel axis}\label{eq:BC-rho-x}\\
\int d\mathbf{r} d\theta \rho(\mathbf{r},\theta)=1
 && \text{\ldots \,normalization of probability density}\label{eq:BC-norm}\\
  J_\theta|_{\theta=0}=J_\theta|_{\theta=2\pi} && \text{\ldots \,periodicity of $J_\theta$} \label{eq:BC-period-theta}\\
 \int_0^{2\pi} d\theta\int_{\frac{-h(x)+R}{d}}^{\frac{h(x)-R}{d}} dz J_\theta(\mathbf{r},\theta)=0 && \text{\ldots \, no net rotational flux\,.}\label{eq:BC-Jtheta}
\end{eqnarray}
where $L$ is the length of the channel. 
We remark that Eq.~(\ref{eq:BC-Jtheta}) stems from the 
mirror symmetry about the channel axis, which implies
$\rho(x,z,\theta)=\rho(x,-z,\pi-\theta)$ and $\omega(x,z,\theta)=-\omega(x,-z,\pi-\theta)$.
This implies that the integrated rotational flux as formulated in Eq.\ (\ref{eq:BC-Jtheta}) vanishes
for all positions along the channel 
axis
hence preventing the onset of local recirculation of microswimmers.

Equation~(\ref{eq:smol_7}) is quite involved because it accounts for contributions stemming from both the translational 
($\mathbf{J}_{\mathbf{r}}$) and rotational ($J_{\theta}$) 
fluxes.
In order to disentangle these contributions, we will follow Ref.~\cite{Elgeti2013} and assume $\mathrm{Pe}_{t} /  \mathrm{Pe}_{r} \ll  1$. 
We remark that
for the thermal or Stokes-Einstein values of
the diffusion coefficients $D_t$ and $D_r$
the condition $\mathrm{Pe}_{t} /  \mathrm{Pe}_{r} = 3d^2 / 4R^2  \ll  1$ 
gives
$d^2 \ll R^2$, 
which 
implies narrow channels. 
The ratio $\mathrm{Pe}_{t} / \mathrm{Pe}_{r}$
can be affected by the activity of the microswimmers. 
For example,
for a 
self-diffusiophoretic particle
with $R=1\mu\text{m}$
the rotational decorrelation time has been measured~\cite{Golestanian2007} to be 
$D^{-1}_r\sim 4\, \text{sec}$.
Hence,
the condition $\mathrm{Pe}_{t} /  \mathrm{Pe}_{r}=d^2 D_r/D_t\ll 1$
together 
with $D_t=k_BT/(6\pi\eta R)$  
gives
an upper bound $d\simeq 1 \mu\text{m}$, i.e., the channel 
half width
is twice the particle
radius.
Finally, we note
that the regime $d^2\ll R^2$ can be attained for all synthetic and biological swimmers by properly tuning the 
accessible
average width of the channel,
$2d$.
We now formally solve Eq.\ (\ref{eq:smol_7}) for the two cases of plane and corrugated channel walls using an expansion in
$\mathrm{Pe}_{t} /  \mathrm{Pe}_{r} \ll  1$.

\subsection{Plane channel walls}
For a channel with plane 
walls, $h(x)=h_{0}$, the probability density $\rho$ and the velocities $v$, $\omega$ are uniform
along $x$. 
In the absence of any spontaneous symmetry breaking along the $x$ axis, this gives $J_x=0$. The boundary condition, Eq.~(\ref{eq:BC-Jz}), reads:
\begin{eqnarray}
 J_z(z=\pm 1,\theta) &=& 0 \, .
 \label{eq:BC-rho-flat-z}
\end{eqnarray}
In zeroth order in $\mathrm{Pe}_{t} / \mathrm{Pe}_{r}$ the rotational contribution in Eq.~(\ref{eq:smol_7}) is negligible and we have:
\begin{equation}
\nabla_{\mathbf{r}}^{2}\rho +\nabla_{\mathbf{r}}\cdot\left[\rho\nabla_{\mathbf{r}}\beta C\right]
-\mathrm{Pe}_{t}\nabla_{\mathbf{r}}\cdot\left[\left(\mathbf{n}+\mathbf{v}\right)\rho\right] =0
\label{eq:flat-rho-z}
\end{equation}
Since $\rho$ does not depend on $x$, which means zero flow along $x$ ($J_x=0$), Eq.~(\ref{eq:flat-rho-z}) can be integrated once and reduces to
\begin{equation}
\frac{\partial}{\partial z}\rho(z,\theta)-
\left(    \mathrm{Pe}_{t} 
[\cos\theta+v_{z}(z,\theta)]+\frac{\partial}{\partial z} \beta C(z) \right) \rho(z,\theta)=-J_{z,0}(\theta) 
 = 0 \, .   
\label{eq:smol_zeroth}
\end{equation}
The constant probability flux $J_{z,0}(\theta)$ vanishes due to Eq.~(\ref{eq:BC-rho-flat-z}).  Equation~(\ref{eq:smol_zeroth})
can be integrated and one obtains
\begin{equation}
\rho_{0}(z,\theta)=\lambda_{0}(\theta)   e^{\mathrm{Pe}_{t}\left[z\cos\theta+\int v_{z}(z,\theta)dz \right]  +\beta C(z)}   
= \lambda_{0}(\theta) e^{-\mathrm{Pe}_{t} W(z,\theta)}      
\label{eq:smol_sol_1}
\end{equation}
In the last equation we have introduced the effective potential
\begin{equation}
W(z,\theta)=\begin{cases}
-z\cos\theta-\int v_{z}(z,\theta)dz & |z|\leq 1\\
\infty & |z|>1
\label{eq:def-W}
\end{cases}
\end{equation}
using the properties of the hard-core potential $C(z)$ from Eq.\ (\ref{eq:def_C}).
We recall that all lengths are measured in units of $d=h-R$.
The factor $\lambda_0$ is determined by requiring that $J_\theta$, once calculated using Eq.~(\ref{eq:smol_sol_1}), 
fulfills the local isotropy of Eq.~(\ref{eq:BC-Jtheta}). 
In particular, if $J_x=0$, Eq.~(\ref{eq:BC-Jtheta}) reduces to\footnote{
Equation (\ref{eq:new-bnd-cond}) is obtained 
by integrating
the steady-state condition along $z$, namely 
$\int dz (\partial_x J_x + \partial_z J_z+\partial_\theta J_\theta)=0$.
Then,
using  Eq.~(\ref{eq:BC-rho-flat-z}) and the fact that for plane channel walls we have $J_x=0$, 
this condition reduces to
$\int dz \partial_\theta J_\theta=\partial_\theta \int dz  J_\theta=0$,
which
 implies that $\int dz  J_\theta$ is a constant. Finally, due to Eq.~(\ref{eq:BC-Jtheta}) we 
 obtain
 $\int dz  J_\theta=0$.}
\begin{equation}
 \int dz J_\theta(z,\theta)=0
 \label{eq:new-bnd-cond}
\end{equation}
Rewriting $\rho_{0}=\lambda_{0}\hat{\rho}_{0}$, using $J_\theta$ from Eq.~(\ref{eq:def-ang-flux}), and Eq.~(\ref{eq:new-bnd-cond}) we obtain:
\begin{equation}
\frac{1}{D_r} \int_{-\infty}^{\infty} J_{\theta}(z,\theta)dz=-\lambda_{0}\frac{\partial}{\partial \theta}\int_{-\infty}^{\infty}\hat{\rho}_{0}
dz-\int_{-\infty}^{\infty}\hat{\rho}_{0}dz\frac{\partial}{\partial \theta}\lambda_{0}+\frac{\omega_{0}}{D_{r}}
\lambda_{0}\int_{-\infty}^{\infty}\omega\hat{\rho}_{0}dz=0
\end{equation} 
whose solution reads: 
\begin{equation}
\lambda_{0}(\theta)=\bar{\rho}_{0}\dfrac{1}{\int_{-\infty}^{\infty}\hat{\rho}_{0}dz}\exp\left[\dfrac{\omega_{0}}{D_{r}}\int d\theta \dfrac{\int_{-\infty}^{\infty}\omega(z,\theta)\hat{\rho}_{0}(z,\theta)dz}{\int_{-\infty}^{\infty}\hat{\rho}_{0}(z,\theta)dz}\right]
\end{equation}
Finally, we 
substitute the last expression in Eq.~(\ref{eq:smol_sol_1}) and write
\begin{equation}
\rho_{0}(z,\theta)=\bar{\rho}_{0}\dfrac{e^{- \mathrm{Pe}_{t}  W(z,\theta)}}{\int_{-\infty}^{\infty}e^{-  \mathrm{Pe}_{t}  W(z,\theta)}dz}
\exp\left[\frac{4}{3}\frac{R}{d}  \mathrm{Pe}_t   \int d \theta\frac{\int_{-\infty}^{\infty}\omega(z,\theta)
e^{- \mathrm{Pe}_{t}  W(z,\theta)}dz} {\int_{-\infty}^{\infty}e^{-  \mathrm{Pe}_{t}  W(z,\theta)}dz}\right] \, ,
\label{eq:smol_zeroth_final_sol}
\end{equation}
where $\bar{\rho}_{0}$ is determined by the normalization condition $\int_{-\pi}^{\pi}d\theta\int_{-\infty}^{\infty}\rho_{0}(z,\theta)dz=1$. 
We also rewrote $\omega_0/D_r=4/3 v_0 R/D_t=4/3  \mathrm{Pe}_t  R/d$ using the thermal values for $D_r$ and $D_t$.
We remark that neglecting the hydrodynamic coupling with the walls, namely for $v=\omega=0$, Eq.~(\ref{eq:smol_zeroth_final_sol}) reduces to Eq.~(9) of Ref.~\cite{Elgeti2013}. Interestingly,  the probability distribution of Eq.\ (\ref{eq:smol_zeroth_final_sol}) does not factorize in the two ``natural'' effective potentials, namely $W(z,\theta)$ and $\int \omega(z,\theta)d\theta $,
from which the stall force and torque, necessary to stop the microswimmer, can be derived.
This implies that the translational and rotational degrees of freedom are quite entangled. In order to get insight into their relative contributions to the particle probability distribution, we expand Eq.~(\ref{eq:smol_zeroth_final_sol})  
for small values of $\mathrm{Pe}_t$ and for $|z|<1$. In particular, at first order in $\mathrm{Pe}_t$ we obtain:
\begin{equation}
 \dfrac{\rho_{0,1}(z,\theta)}{ \bar{\rho}_0 }\simeq 
\dfrac{1}{2}
 \left[ 1-  \mathrm{Pe}_t 
 \Big\{  W(z,\theta) + \dfrac{1}{2} \int_{-1}^{1} W(z,\theta)dz
+\frac{2}{3}\frac{R}{d} \int d\theta \int_{-1}^{1}\omega(z,\theta)dz
\Big\} \right]
 +\mathcal{O}( \mathrm{Pe}_t^2 ) \, .
\label{eq:def-rho-01}
\end{equation}
Note that the angular velocity only contributes with its mean value and this term vanishes for pure source-dipole swimmers [see Eq.~(\ref{eq:omega-flat}), which shows that $\omega$ is an odd function in $z$.]

\subsection{Corrugated channel walls}

In order to characterize the dynamics along the channel axis, we determine and discuss the probability density $p(x)$ to find the microswimmer at position $x$ for any $\theta$, $z$ and also the mean drift velocity in $x$ direction. 
To do so, we take inspiration from the Fick--Jacobs approximation and extend it to the case of active particles. Accordingly, we start with an ansatz for the full probability density:
\begin{equation}
\rho(x,z,\theta)=p(x)\dfrac{g(x,z,\theta)}{\int_{-\infty}^{\infty}\int_{-\pi}^{\pi}g(x,z,\theta)dzd\theta}
\label{eq:anzatz-corr}
\end{equation}
where 
\begin{equation}
g(x,z,\theta)=\dfrac{e^{-\mathrm{Pe}_tW(x,z,\theta)}}{\int_{-\infty}^{\infty}e^{-\mathrm{Pe}_t W(x,z,\theta)}dz}
\exp\left[ \dfrac{4}{3}\dfrac{R}{d}\mathrm{Pe}_t\int d\theta
\frac{\int_{-\infty}^{\infty}\omega(x,z,\theta)e^{-\mathrm{Pe}_tW(x,z,\theta)}dz}
{\int_{-\infty}^{\infty}e^{-\mathrm{Pe}_tW(x,z,\theta)}dz}
\right]
\label{eq:g-corrugated}
\end{equation}
has the same functional form as in Eq.~(\ref{eq:smol_zeroth_final_sol}) for plane channel walls and thus fulfills the boundary conditions 
of Eqs.~(\ref{eq:BC-period-theta}), (\ref{eq:BC-Jtheta}), and (\ref{eq:BC-Jz}) (for the latter see Appendix A of Ref.\cite{Marconi2015}).
In making this ansatz we are again in the limit of $\mathrm{Pe}_{t} / \mathrm{Pe}_{r} \ll 1$ and also assume small variations of the channel width, $|\partial h(x) / \partial x| \ll 1$. We also note that $\omega$ and $v$ account for the local slope of the channel walls and thus they now depend on $x$ (see Appendix~\ref{app:3}). 
This ansatz for $\rho(x,z,\theta)$ in Eqs.~(\ref{eq:anzatz-corr}) and (\ref{eq:g-corrugated}) represents the extension of the Fick--Jacobs approximation to active particles.

We integrate Eq.\ (\ref{eq:smol_7}) along $z$ and $\theta$ and use the boundary conditions from Eqs.~(\ref{eq:BC-Jz}),(\ref{eq:BC-Jtheta}) together with $|\partial_x h(x)|\ll 1$~\cite{Marconi2015} to obtain
\begin{equation}
\int_{-\infty}^{\infty}\int_{-\pi}^{\pi} \left\{
\frac{\partial^{2}}{\partial x^2}\rho+\frac{\partial}{\partial x}\left[\rho\frac{\partial}{\partial x}\beta C\right]-\mathrm{Pe}_t\frac{\partial}{\partial x}\left[\left(\sin\theta+v_x\right)\rho\right]  \right\}
dzd\theta=0 \, ,
\label{eq:smol-1-corr}
\end{equation} 
or, equivalently, after integrating once:
\begin{equation}
\int_{-\infty}^{\infty}\int_{-\pi}^{\pi}\left\{\frac{\partial}{\partial x}\rho+\rho \left[\frac{\partial}{\partial x}\beta C-\mathrm{Pe}_t
\left[\left(\sin(\theta)+v_{x}\right)\right]\right]\right\}  dzd\theta= - I_{x} \, ,
\label{eq:smol_ch_1-1}
\end{equation}
where $I_x$ is the probability current along the $x$ direction. We use the ansatz from Eq. (\ref{eq:anzatz-corr}) to show that
\begin{equation}
\int_{-\infty}^{\infty}\int_{-\pi}^{\pi}\frac{\partial}{\partial x}\rho(x,z,\theta)dzd\theta=\frac{\partial}{\partial x}p(x)
\label{eq:def-dervi-p}
\end{equation}
and also define the average drift velocity $v_D(x)$ at axial position $x$,
\begin{equation}
v_D(x) = -
\beta\frac{\partial}{\partial x}\chi(x)= 
\int_{-\infty}^{\infty}\int_{-\pi}^{\pi}
\left[-\frac{\partial}{\partial x}\beta C+
\mathrm{Pe}_t\left(\sin(\theta)+v_{x}\right)\right]\dfrac{g(x,z,\theta)}{\int_{-\infty}^{\infty}\int_{-\pi}^{\pi}g(x,z,\theta)dzd\theta} dzd\theta\, ,
\label{eq.vxmean}
\end{equation}
as the derivative of the effective total
potential $\chi(x)$, which in one dimension can always be defined.
Using Eqs.~(\ref{eq:def-dervi-p}) and (\ref{eq.vxmean}), we are able to write Eq.~(\ref{eq:smol_ch_1-1}) in compact
form,
\begin{equation}
\frac{\partial}{\partial x}p(x)+\beta  p(x)\frac{\partial}{\partial x}\chi(x)=-I_{x} \, ,
\label{eq:smol_ch_sol}
\end{equation}
the solution of which reads
\begin{equation}
p(x)=e^{-\beta\chi(x)}\left[-I_{x}\int_0^x e^{\beta\chi(x')}dx'+\Theta\right] \, .
\label{eq:prob-corr}
\end{equation}
Here, $\Theta$ and $I_{x}$ are determined by imposing the boundary condition of Eq.~(\ref{eq:BC-rho-x}), and the normalization $\int p(x) dx =1$.
For $I_x=0$, Eq.~(\ref{eq:prob-corr}) reduces to
\begin{equation}
p(x)=\Theta e^{-\beta \chi(x)} \, .
\label{eq:prob-corr-red}
\end{equation}
Finally, we can use the current $I_x$  to define the net drift velocity along the $x$ axis as
\begin{equation}
 v_d=I_x L \, .
\label{eq.vd}
\end{equation}
This completes the full solution of our problem. 
With the help of $\rho(x,z,\theta)$ from Eqs.~(\ref{eq:anzatz-corr}) and (\ref{eq:g-corrugated}), we can calculate $v_D(x)$ from Eq.~(\ref{eq.vxmean}), integrate it once to obtain $\chi(x)$, and then determine the probability density $p(x)$ from Eq.~(\ref{eq:prob-corr}).

Even though the functional form of Eq.~(\ref{eq:prob-corr}) is similar to that obtained for passive systems governed solely by 
conservative forces~\cite{Reguera2001}, the difference between Eq.~(\ref{eq:prob-corr}) and the corresponding expression in 
the passive case lies in the form of the effective total potential, $\chi(x)$, the derivative of which can be expressed as
\begin{equation}
-\beta\frac{\partial}{\partial x}\chi(x)
=
\mathrm{Pe}_t\int_{-\infty}^{\infty}\int_{-\pi}^{\pi}\dfrac{\left(\sin(\theta)+v_{x}\right)g(x,z,\theta)}{\int_{-\infty}^{\infty}\int_{-\pi}^{\pi}g(x,z,\theta)dzd\theta} dzd\theta 
-\beta \frac{\partial}{\partial x}\mathcal{A}(x) \, .
\label{eq:vel-x-corr}
\end{equation}
Here, we have introduced the gradient of an effective entropic potential $\mathcal{A}(x)$, which generalizes the entropic potential of passive systems~\cite{Reguera2001} and derives from
\begin{equation}
 \beta \frac{\partial}{\partial x}\mathcal{A}(x)=
 \beta \int_{-\infty}^{\infty}\int_{-\pi}^{\pi} 
  \left[ \frac{\partial}{\partial x} C(x,z) \right]   
\dfrac{ g(x,z,\theta)}{\int_{-\infty}^{\infty}\int_{-\pi}^{\pi}g(x,z,\theta)dzd\theta} dzd\theta =-2 
  \left[  \frac{\partial}{\partial x}h(x)   \right]  
\int_{-\pi}^{\pi}\dfrac{g(x,h(x)-R,\theta)}{\int_{-\infty}^{\infty}\int_{-\pi}^{\pi}g(x,z,\theta)dzd\theta} d\theta \, .
 \label{eq:def-A}
\end{equation}
To arrive at the last expression, we used the explicit form of $C(x,z)$ from Eq.~(\ref{eq:def_C}) 
(see Appendix~\ref{app:A} for details).
In particular, for 
passive systems with only conservative interactions with bounding walls, $\mathrm{Pe}_t=0$ and from Eq.~(\ref{eq:smol_ch_1-1}) one has $g(x,z,\theta) \propto\exp[-\beta C(x,z)]$.
Accordingly, the first term on the right-hand side of Eq.~(\ref{eq:vel-x-corr}) vanishes and Eq.~(\ref{eq:def-A}) can be integrated leading to
\begin{equation}
\beta \chi(x)=\beta\mathcal{A}_{o}(x)=-\ln \left[\int_{-\infty}^{\infty}\int_{-\pi}^{\pi}\exp[-\beta C(x,z)] dzd\theta\right]=-\ln\left[\frac{h(x)-R}{d}\right] \, .
 \label{eq:def-Ao}
\end{equation}
Therefore, for passive systems, $\mathcal{A}(x)$ reduces to the usual entropic potential\footnote{The term ``entropic'' stems from the fact that $h(x)-R$ measures the number of states available in the transverse direction and therefore $\ln[(h(x)-R)/d]$ is proportional to the local entropy.} of Eq.~(\ref{eq:def-Ao}).
In contrast, in the present  active case 
$\mathrm{Pe}_t\neq 0$ and the effective interactions are not conservative. Accordingly, $g(x,z,\theta) \nsim \exp[-\beta C(x,z)]$, and 
the form of both $\chi(x)$ and $\mathcal{A}(x)$ is more involved.
Interestingly, Eq.~(\ref{eq:vel-x-corr}) identifies two separate contributions to the mean drift velocity $v_D(x)$, namely an ``active'' one [the first term on the rhs of Eq.~(\ref{eq:vel-x-corr})] and an ``entropic'' one [the second term on the rhs of Eq.~(\ref{eq:vel-x-corr})]. Therefore, by tuning the geometry of the channel and the activity of the particles via  
the P\'eclet number, it is possible to tune  the relevance of the entropic against active contribution.
Moreover, the sign of the active contribution depends on the hydrodynamic coupling to the bounding walls and hence on the characteristic flow profile induced by the swimmer. Accordingly, for those cases for which the entropic and the active contribution have opposite sign, Eq.~(\ref{eq:vel-x-corr}) predicts the existence of a critical value of the P\'eclet number, $\mathrm{Pe}_c$, for which the dynamics crosses over from 
entropy-controlled (small $\mathrm{Pe}_t$) to activity-controlled (large $\mathrm{Pe}_t$). In order to gain some insight into the magnitude of $\mathrm{Pe}_c$, we assume 
\begin{equation}
 g(x,z,\theta)\simeq\bar{g}(x,z)=g_0e^{-\beta C(x,z)} \, ,
 \label{eq:g-active-ansatz}
\end{equation}
namely the density is constant inside the channel and zero outside\footnote{The probability distribution in Eq.~(\ref{eq:g-active-ansatz}) is the one that should be expected for passive systems, i.e., for $\mathrm{Pe}_t=0$. Here,
we assume that Eq.~(\ref{eq:g-active-ansatz}) roughly captures the functional form of $g(x,z,\theta)$ also for $\mathrm{Pe}_t\neq0$. As discussed in the Results section, such an assumption is justified {\it a posteriori} due to the good matching of the predictions that can be derived, Eq.~(\ref{eq:Pe-C-3}), when compared with the numerical integration of $g(x,z,\theta)$.}. In this case, using Eq.~(\ref{eq:def_C}), Eq.~(\ref{eq:vel-x-corr}) reads:
\begin{equation}
v_D(x) =
 \mathrm{Pe}_t \frac{
\left<v_x\right> 
 }{4\pi h(x)} +  \frac{\partial}{\partial x} \ln\left[ 2h(x)\right]
 \label{eq:Pe-C-1}
\end{equation}
where $\left<v_x\right>=\int_{-\infty}^{\infty}\int_{-\pi}^{\pi}v_x\bar{g}dzd\theta$ and
$\int_{-\infty}^{\infty}\int_{-\pi}^{\pi}\sin(\theta)\bar{g}dzd\theta=0$ due to the symmetry of $\bar{g}$.
Setting 
$v_D(x) = 0$
in Eq.~(\ref{eq:Pe-C-1}), we identify the cross-over value of the P\'eclet number, 
\begin{equation}
 \mathrm{Pe}_c =
 \left|\frac{4\pi\partial_x h(x)}{\left\langle v_{x}\right\rangle}\right|
 \simeq \frac{4\pi}{| \left\langle v_{x}\right\rangle|}\frac{2 h_1}{L/2} \, ,
 \label{eq:Pe-C-2}
\end{equation}
where in the last step we used the amplitude $h_1$ of the corrugation of the channel walls and we made the approximation $\partial_x h(x)\simeq \frac{2 h_1}{L/2}$. 
Moreover, assuming that the magnitude of $\left\langle v_{x}\right\rangle $ is comparable to $v_0$, i.e., for $\left\langle v_{x}\right\rangle \simeq 1$ we obtain 
\begin{equation}
 \mathrm{Pe}_c\simeq 16\pi\frac{h_1}{L} \, .
 \label{eq:Pe-C-3}
\end{equation}

\section{Results}

As shown by Eqs.~(\ref{eq:smol_zeroth_final_sol}) and (\ref{eq:prob-corr}), the dynamics of an active particle inside a channel is strongly affected by the hydrodynamic coupling to the bounding walls. 
Since an exact solution of the Stokes equation for such a geometry is lacking, we have to rely on approximate
solutions \pa{or generic models}. In particular, \pa{in order to derive general results that do not depend on a specific swimming mechanism, we study model microswimmers such as pushers, pullers, and source dipoles, the velocity fields of which correspond to 
the lowest-order contributions in a far--field multipole expansion of real microswimmers. 
Moreover, we neglect that their finite size disturbs the fluid velocity
field of the generic swimmers as it was done in other studies~\cite{Ortiz2005,LaugaPRL2008,Baskaran15092009,Spagnolie2012,Lauga2014,Hennes2014,Schaar2015}.
For such model microswimmers the effective interactions with the channel walls can be captured by the method of
images\footnote{\pa{In order to keep our treatment simple, we have truncated the set of images at the first iteration, see Appendix~\ref{app:3}.}}. Interestingly, it has been shown that for some cases~\cite{Spagnolie2012} 
these generic far--field expansions quantitatively capture the hydrodynamic interactions with bounding walls even when the microswimmers are very close to these walls.
As a result, hydrodynamic interactions} \pa{between model microswimmers and channel walls are accounted for by the linear and angular velocities induced by the image flow fields at the center of mass of the microswimmers. Finally, the fact that real microswimmers have a finite extent and in order to avoid divergences in the hydrodynamic coupling, we prevent model microswimmers from approaching channel walls not closer than their ``radius'' $R$.
}

In this perspective, we exploit \pa{generic} far--field approximations to grasp the relevance of the swimmer type and mechanism, and we characterize the effective dynamics of four different kinds of swimmers~\cite{Spagnolie2012}. In particular, we consider swimmers, which generate either a source dipole of strength $q$ or a force dipole of strength $p$, distinguishing pushers ($p>0$) and pullers ($p<0$), and compare them to pure active Brownian particles, which do not hydrodynamically couple to the bounding walls. The different swimmer types are determined by the wall-induced linear ($\mathbf{v}$) and angular ($\omega$) velocity fields, for which we take the far--field expressions as summarized in Appendix \ref{app:3}. 
Source dipoles also called ``neutral swimmers'', 
the flow fields of which decay as $1/r^3$,
resemble the motion of biological microswimmers such as \textit{Paramecia}~\cite{Ishikawa2006} or artificial swimmers as the active emulsion droplets of Refs. \cite{Thutupalli2011,Schmitt2013}.
Typical force dipoles in nature are bacteria such as \textit{E.Coli} that are characterized by $p>0$ and are named ``pushers'' since they push the fluid along the direction of motion, whereas algae like \textit{Clamydomonas} are characterized by $p<0$ and are named ``pullers'' since they pull on the fluid along the direction of motion.
Their flow fields always decay as $1/r^2$.
In the following we characterize the steady-state probability distribution along a microchannel with periodic boundary conditions at both channel ends.

\subsection*{Plane channel walls}

\begin{figure}
 \includegraphics[scale=0.42]{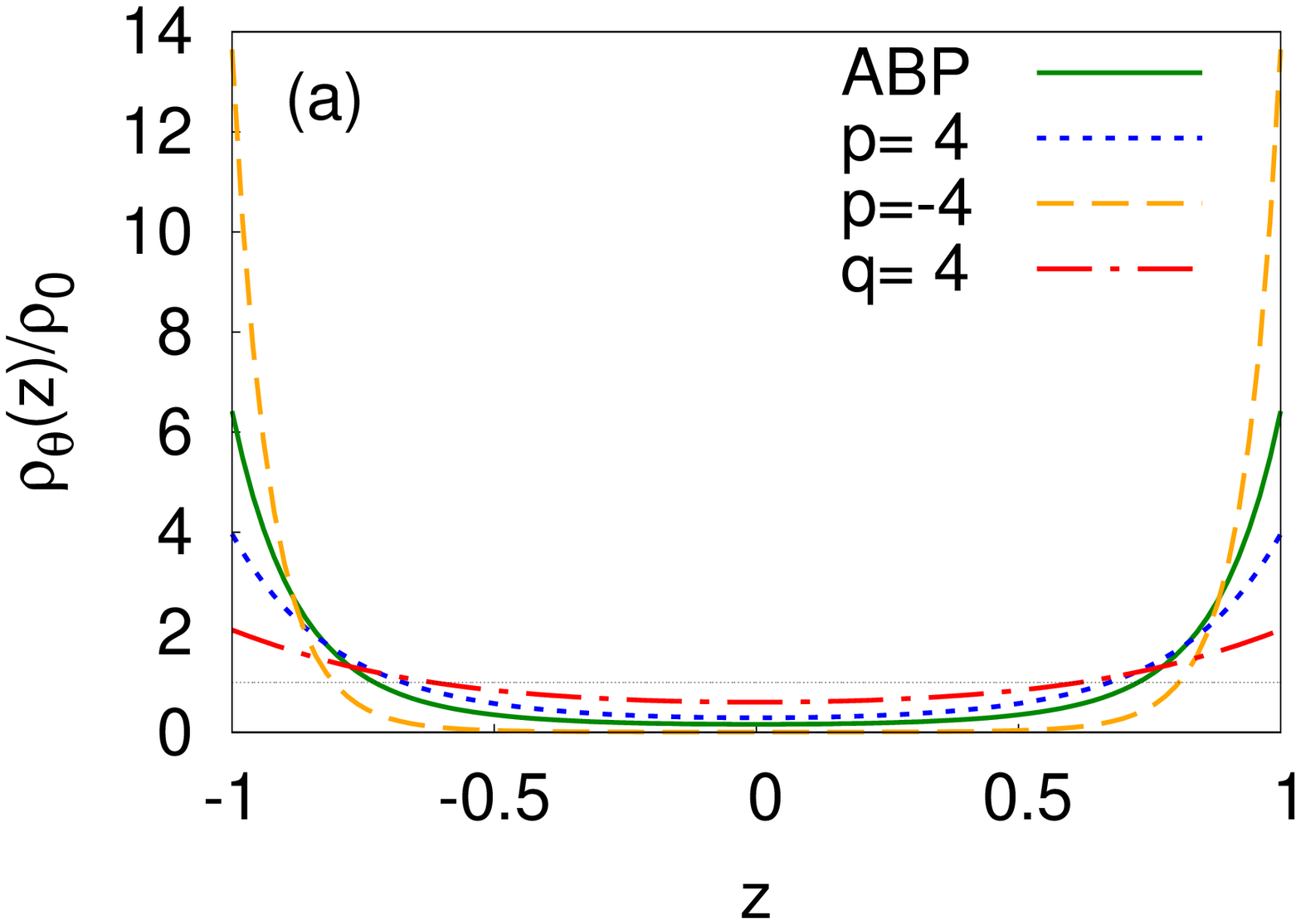}\,\,\,\,\,\,\includegraphics[scale=0.42]{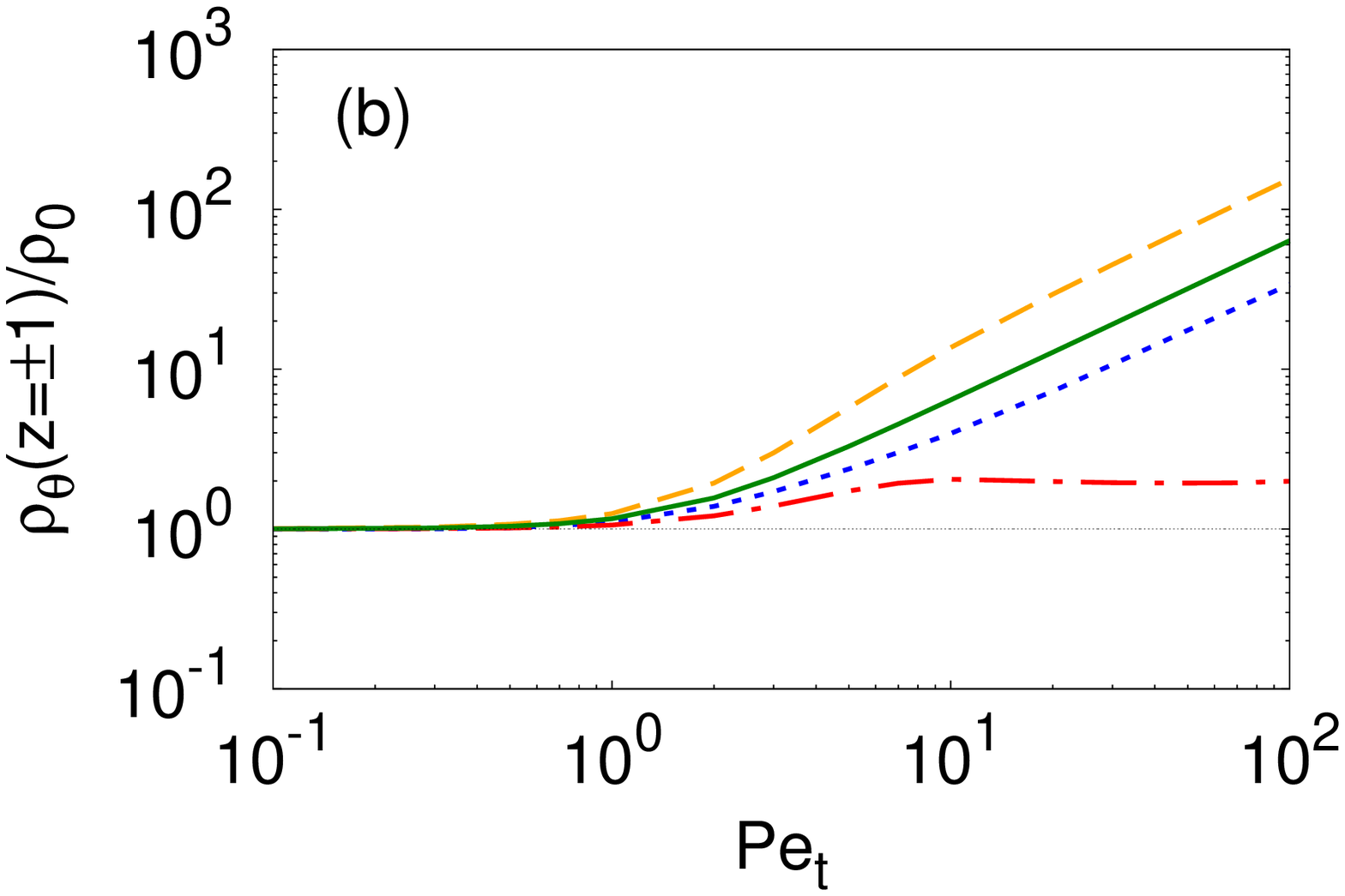}\\
 \includegraphics[scale=0.42]{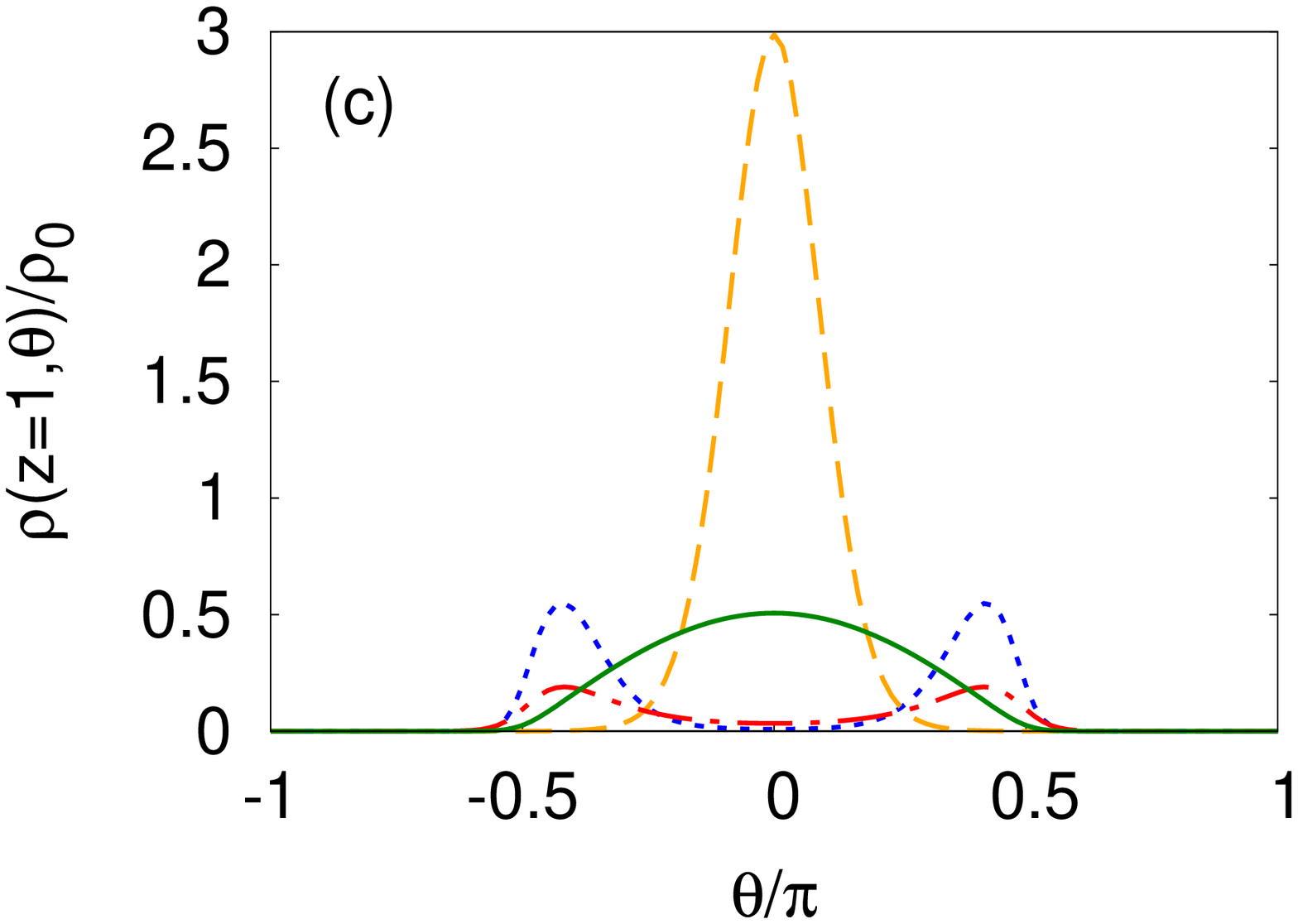}\,\,\,\,\,\,\includegraphics[scale=0.42]{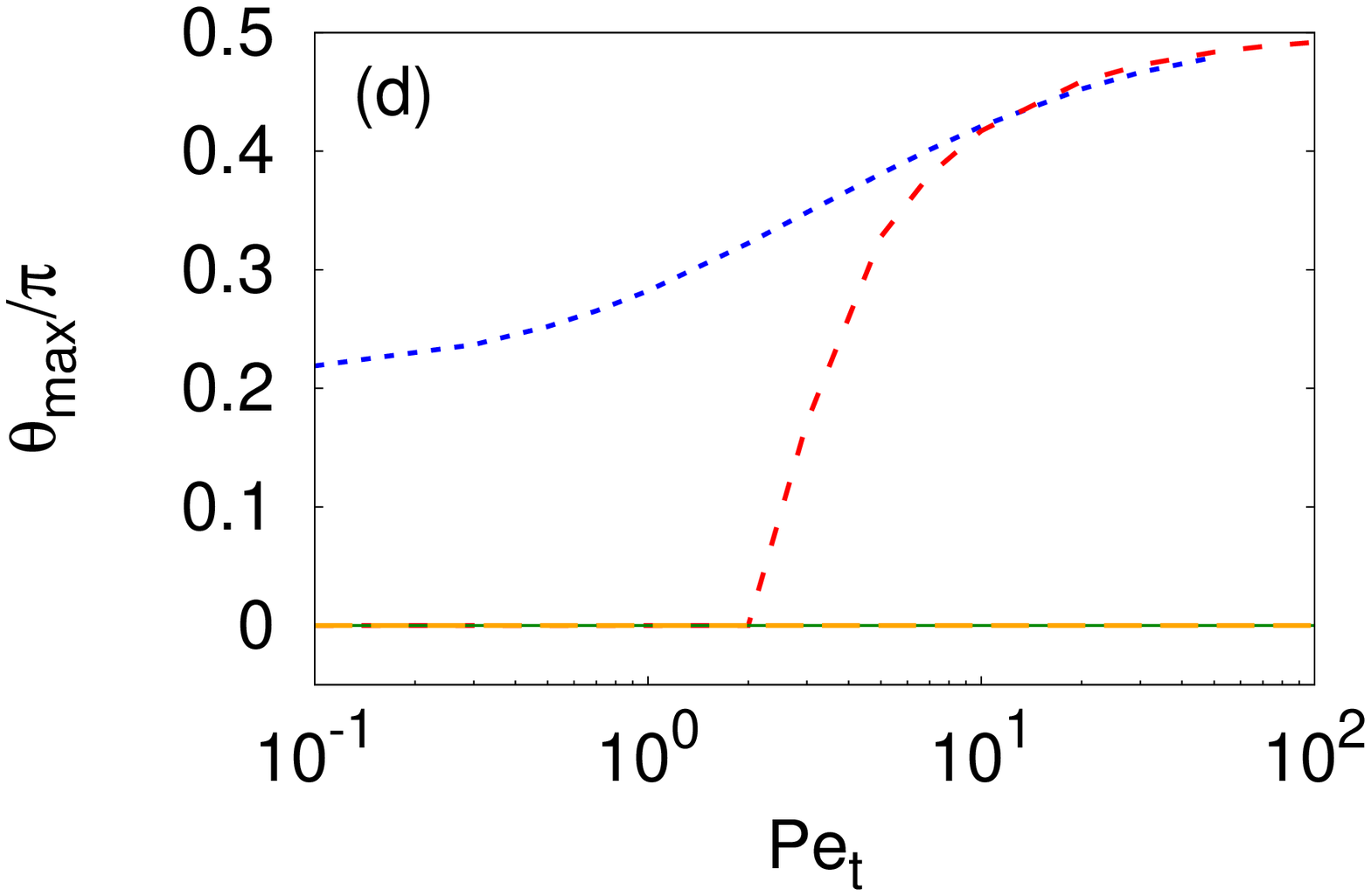}
 \caption{Plane channel walls. a)
Lateral density profile $\rho_\theta(z)$, normalized by the density of a passive particle, $\rho_0 = 1/(2h_0)$, as a function of the lateral position $z$ for pushers ($p=4$, blue dashed), pullers ($p=-4$, orange long-dashed), and  source dipoles ($q=4$, red dot-dashed) embedded in a channel with half width 
$h_0=3R/2$ and $d=h_0-R=R/2$, where $d$ is the space available to the center of mass of the active particle along the lateral 
direction. We recall that $d$ is the unit length of our model.
The case of an active Brownian particle (ABP) is reported as a reference (solid green). All swimmers are characterized by 
$\mathrm{Pe}_t=10$. (The color scheme and parameter values are kept for the following plots unless otherwise stated.)
b) 
Lateral density at contact with the wall, $\rho_\theta(z = \pm 1)$, as a function of the translational P\'eclet number $\mathrm{Pe}_t$. 
c) Density at contact with the wall $\rho(z=\pm 1,\theta)$ as a function of the orientation angle $\theta$. 
d) Orientation angle $\theta_{max}$ that maximizes the probability at contact, $\rho(z=\pm 1,\theta_{max})$,
as a function of $\mathrm{Pe}_t$.
} 
\label{fig:sm-rho-z}
\end{figure}

Substituting the explicit expression for the angular [Eq.~(\ref{eq:omega-flat})] and linear [Eq.~(\ref{eq:vz-flat})] velocity  provided in 
Appendix~B into Eq.~(\ref{eq:smol_zeroth_final_sol}), we can calculate the density profile in steady state.
At first, we focus on the lateral probability density, 
\begin{equation}
\rho_\theta(z)
=\int_{-\pi}^\pi\rho(z,\theta)d\theta \, ,
\end{equation}
which for plane walls does not depend on the axial coordinate $x$.
As shown in Fig.~\ref{fig:sm-rho-z}(a)  (and in agreement with Ref.~\cite{Elgeti2013}) an active Brownian particle shows an excess accumulation  at the channel 
walls as compared to a passive particle embedded in the same channel. When an active particle stirs the fluid, the induced hydrodynamic coupling alters the previous picture, as shown in Fig.\ \ref{fig:sm-rho-z}(a). In particular, while pullers [orange long-dashed line in Fig.~\ref{fig:sm-rho-z}(a)] show an enhanced accumulation at channel walls as compared to active Brownian particles, pushers [blue dashed line in Fig.~\ref{fig:sm-rho-z}.(a)] and neutral swimmers [red dot-dashed line in Fig.~\ref{fig:sm-rho-z}.(a)] experience a decreased accumulation at the channel walls.
This is in agreement with the larger detention times of pullers found in Ref.~\cite{Schaar2015} compared to active Brownian particles.
Interestingly, the accumulation of active particles at channel walls, $\rho_\theta(z=1)$, shows a non trivial dependence on $\mathrm{Pe}_t$, as shown in Fig.\ref{fig:sm-rho-z}.(b).  
Looking at the dependence of $\rho_\theta(z=1)$ upon increasing $\mathrm{Pe}_t$, we see that active Brownian particles, pushers, 
and pullers are characterized by a monotonic increase of $\rho_\theta(z=1)$.
In contrast, the density of neutral swimmers  the walls shows a plateau for P\'eclet values larger than $\mathrm{Pe}_t\simeq 10$.

The diverse accumulation at channel walls is related to the different angular orientations  at contact. In fact, as shown in 
Fig.~\ref{fig:sm-rho-z}.(c), active Brownian particles and pullers accumulate at the walls mainly orthogonally pointing towards the wall.
In contrast, pushers and neutral swimmers tend to align parallel with the walls 
(see the peaks of $\rho(z=\pm 1,\theta)$  close to $\theta=\pi/2$).
Finally, while the orientation of active Brownian particles and  pullers does not depend on the value of $\mathrm{Pe}_t$, 
pushers and neutral swimmers show a non trivial dependence on $\mathrm{Pe}_t$ and they tend to fully align parallel to the channel walls for increasing values of $\mathrm{Pe}_t$, as shown in Fig.~\ref{fig:sm-rho-z}.(d).
On the contrary, for $\mathrm{Pe}_t\lesssim 1$, the angle $\theta_{\text{max}}$, where $\rho(z=1,\theta)$ attains its maximum, is captured by Eq.~(\ref{eq:def-rho-01}). 
For the case of a plane channel it reads
\begin{equation}
 \rho_{0,1}(z=1,\theta)=\frac{1}{2}+\left[\left(\frac{1}{4}\frac{h_0\cos2\theta}{(h_0+1)^2}+\frac{3}{8}\frac{h_0(1-3\cos^2\theta)}{h_0^2-1}-\frac{3}{4}\frac{1-3\cos^2\theta}{h_0}\arctanh\left[\frac{1}{h_0}\right]\right)p +\frac{1}{2}\cos\theta-q \frac{h_0\cos\theta}{(1-h_0^2)^2} \right]\mathrm{Pe}_t
 \label{eq:rho-01-flat}
\end{equation}
where we have used Eqs.~(\ref{eq:omega-flat}),(\ref{eq:vz-flat}). In particular, for active Brownian particles ($q=p=0$) Eq.~(\ref{eq:rho-01-flat}) reduces to
\begin{equation}
 \rho^{\text{ABP}}_{0,1}(z=1,\theta)=\frac{1}{2}+\frac{1}{2}\mathrm{Pe}_t\cos\theta \, ,
 \label{eq:rho-01-flat-ABP}
\end{equation}
whereas for neutral swimmers ($q\neq0$, $p=0$) and pushers/pullers ($q=0$, $p\neq0$) Eq.~(\ref{eq:rho-01-flat}) becomes
\begin{eqnarray}
 \rho^{q\neq0,p=0}_{0,1}(z=1,\theta)&=&\frac{1}{2}+\left[\frac{1}{2}-q \frac{h_0}{(1-h_0^2)^2} \right]\mathrm{Pe}_t\cos\theta\label{eq:rho-01-flat-q}\\
 \rho^{q=0,p\neq0}_{0,1}(z=1,\theta)&=&\frac{1}{2}+\left[\left(\frac{1}{4}\frac{h_0\cos2\theta}{(h_0+1)^2}+\frac{3}{8}\frac{h_0(1-3\cos^2\theta)}{h_0^2-1}-\frac{3}{4}\frac{1-3\cos^2\theta}{h_0}\arctanh\left[\frac{1}{h_0}\right]\right)p +\frac{1}{2}\cos\theta \right]\mathrm{Pe}_t\,\,\,\,\,\label{eq:rho-01-flat-p}
\end{eqnarray}
Interestingly, while active Brownian particles always point mainly towards the wall [$\theta_{\text{max}}=0$ for Eq.~(\ref{eq:rho-01-flat-ABP})], the maximum of $\rho_{0,1}(z=1,\theta)$ for neutral and force-dipole swimmers [see Eqs.~(\ref{eq:rho-01-flat-q}) and (\ref{eq:rho-01-flat-p}), respectively] has a more involved parameter dependence.
In particular, for 
$q \frac{h_0}{(1-h_0^2)^2}<\frac{1}{2}$ [valid
for the values of $h_0$, $d$, and $q$ used in Fig.~\ref{fig:sm-rho-z} (d)]
the density in Eq.~(\ref{eq:rho-01-flat-q}) is maximized for $\theta_{\text{max}}=0$, whereas for 
$q \frac{h_0}{(1-h_0^2)^2}>\frac{1}{2}$ one has $\theta_{\text{max}}=\pi$. 
For pushers/pullers the density in Eq.~(\ref{eq:rho-01-flat-p}) has quite an involved dependence on the parameters. In particular, for the geometric values used in Fig.~\ref{fig:sm-rho-z}, {it has its maximum at $\theta_{\text{max}}=0$ for pullers ($p<0$), whereas for pushers ($p>0$) one obtains $\theta_{\text{max}}\simeq \pi/5$, which is the value, $\theta_{\mathrm{max}}$ tends to in Fig.\  \ref{fig:sm-rho-z}(d).
    
\subsection*{Corrugated channel walls}

\begin{figure}
\includegraphics[scale=0.3]{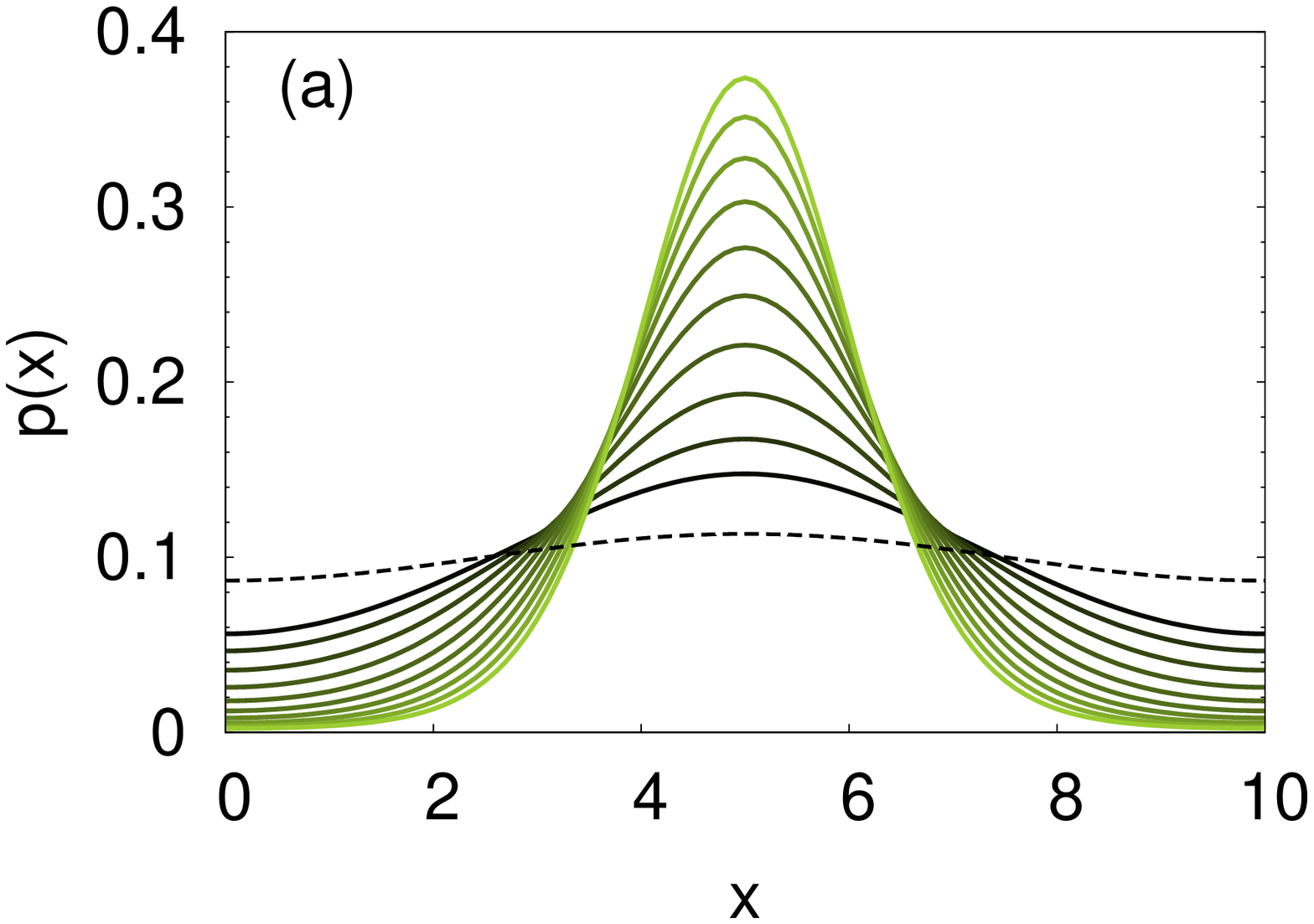} \includegraphics[scale=0.3]{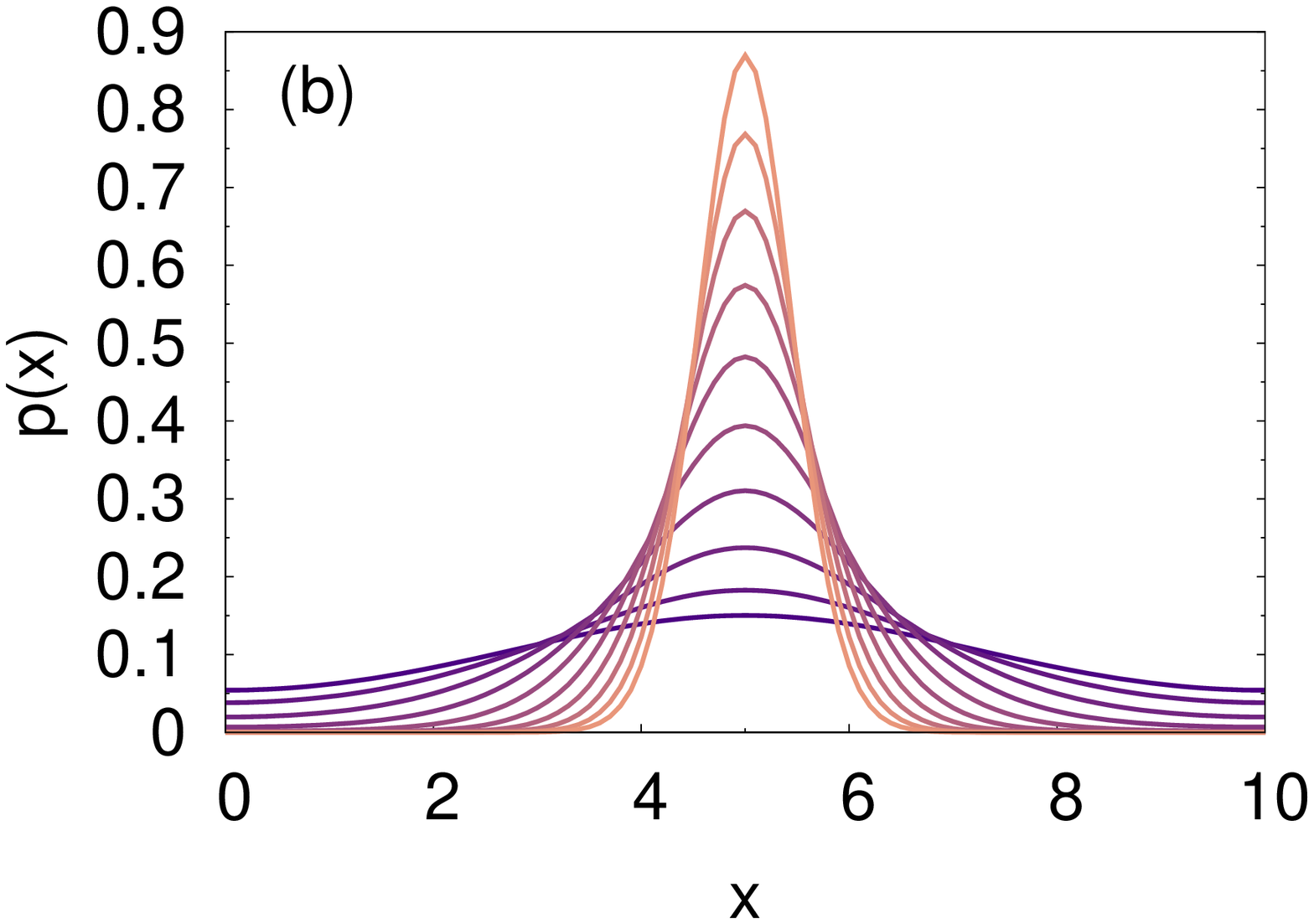}  

\includegraphics[scale=0.3]{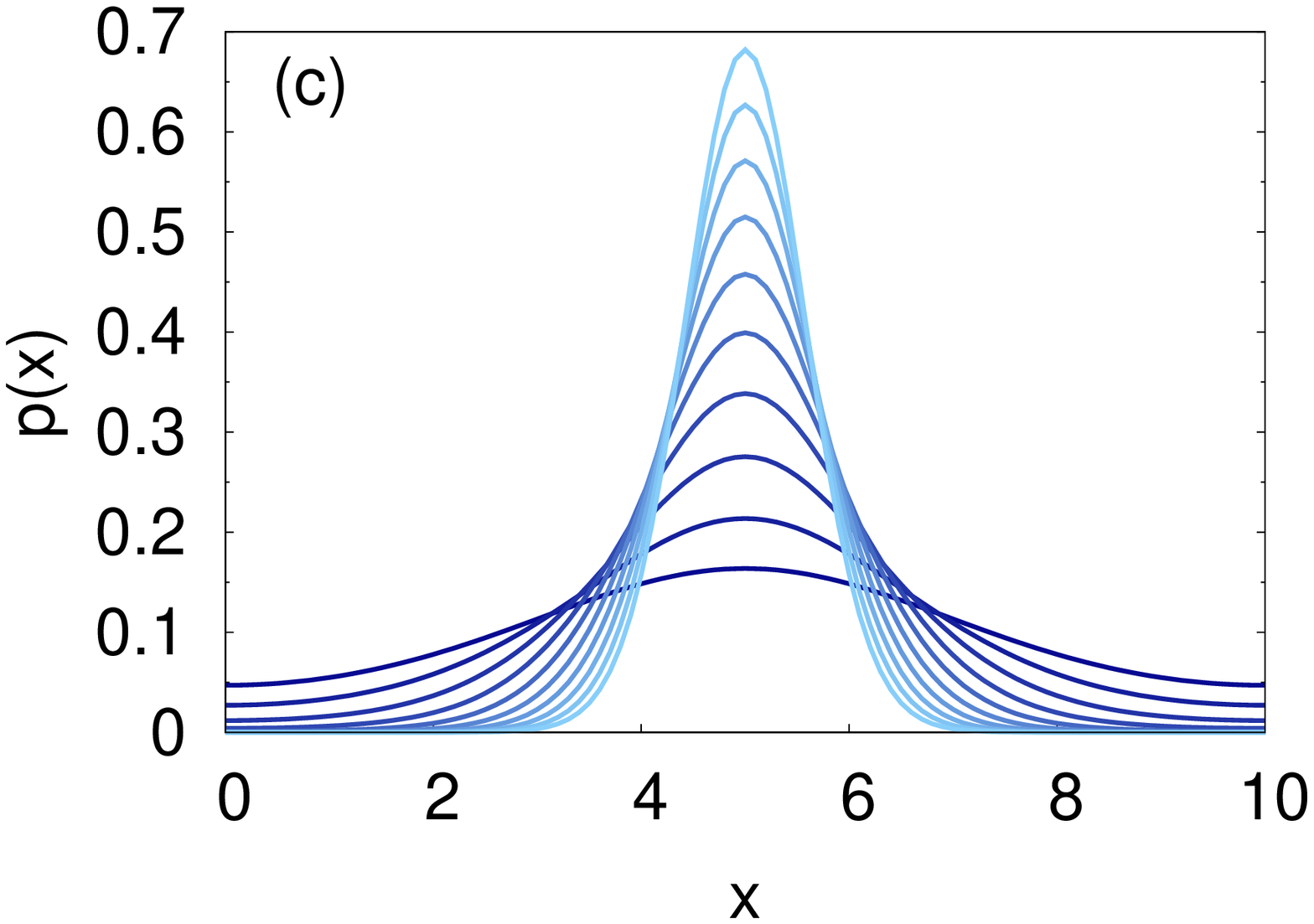} \includegraphics[scale=0.3]{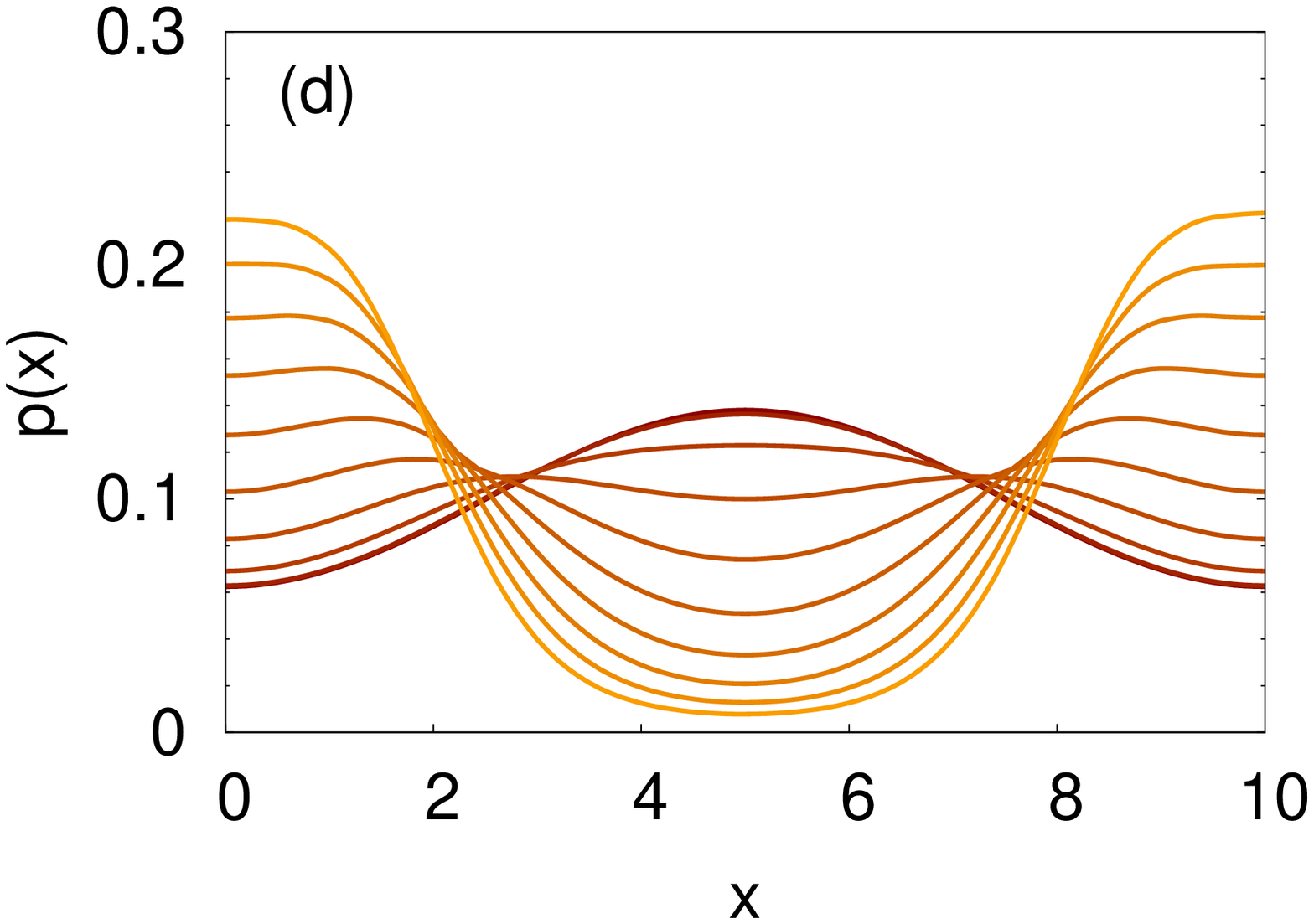}
\caption{Corrugated channel walls: symmetric channel. Probability distribution function $p(x)$ of active Brownian particles (panel a), neutral swimmers ($q=4$, panel b), pushers ($p=4$, panel c), and pullers ($p=-4$, panel d) swimming in a channel characterized by $h_0=3$, $h_1=0.4$, $L=10$, and $L_1=L/2$ (all lengths in units of $d=h_0-R=R/2$).
The different curves correspond to $\mathrm{Pe}_t=1,2,3,4,5,6,7,8,9$, and $10$, where the lines become fainter with increasing $\mathrm{Pe}_t$. For comparison, the profile of $p(x)$ of a passive particle for the same channel shape 
is shown as black dotted line in panel a.}
\label{fig:symm_Pe}
\end{figure}

We have more thoroughly characterized the dynamics of active particles in a channel with corrugated walls, the half width of which is defined as
\begin{eqnarray}
 h(x)=\left\{
\begin{array}{cc}
 h_0-h_1\cos\left[\frac{\pi x}{L_1}\right] & 0\le x<L_1\\
 h_0+h_1\cos\left[\frac{\pi  (x-L_1)}{L-L_1}\right] & L_1\le x \le  L\\
\end{array} \, .
\right.
\label{eq:channel_shape}
\end{eqnarray}
Here, $2h_0$ is the average channel width, $h_1$ the amplitude of the  wall modulation, and $L_1$ characterizes the asymmetry of the wall corrugation along the channel axis (see Fig.~\ref{fig.1}).
In the case of modulated channel walls the explicit expression of the linear ($\mathbf{v}$) and angular ($\omega$) velocity, which we use explicitly in Eqs.~(\ref{eq:anzatz-corr}) and (\ref{eq:g-corrugated}) and all the following implications, must take into account the modulation of the channel width, $2h(x)$. 
If the channel cross section varies on lengths much larger than particle size, $L\gg R$, locally we can regard channel walls as infinite planes tilted by an angle $\psi(x)=\arctan(\partial_x h(x))$ with respect to the channel axis. 
For this case, the hydrodynamic coupling between 
\pa{our model microswimmers} and the tilted channel walls is known in the far--field regime~\cite{Spagnolie2012} (see Appendix \ref{app:3}).
For corrugated channel walls, the mean drift velocity $v_D(x)$ along the channel axis [see Eq.~(\ref{eq:vel-x-corr})] 
is generally non-vanishing and the local drift might result in  a net rectification of particle flow giving a net current $I_x$ or a net drift velocity $v_d = I_x L$, as defined in Eq.\ (\ref{eq.vd}).
Substituting the expressions for the linear velocity along the normal direction
[$v_z(x,z,\theta)$ from Eq.~(\ref{eq:val-sqrm-1-1})], along the longitudinal or axial direction
[$v_x(x,z,\theta)$ from Eq.~(\ref{eq:val-sqrm-1-1x})], and the angular velocity
[$\omega(x,z,\theta)$ from Eq.~(\ref{eq:omega-sqrm-1-1})] into Eqs.~(\ref{eq:def-W}), (\ref{eq:g-corrugated}),
and (\ref{eq.vxmean}), we can calculate the probability distribution along the axial position from Eq.~(\ref{eq:prob-corr}).

\begin{figure}
 \includegraphics[scale=0.35]{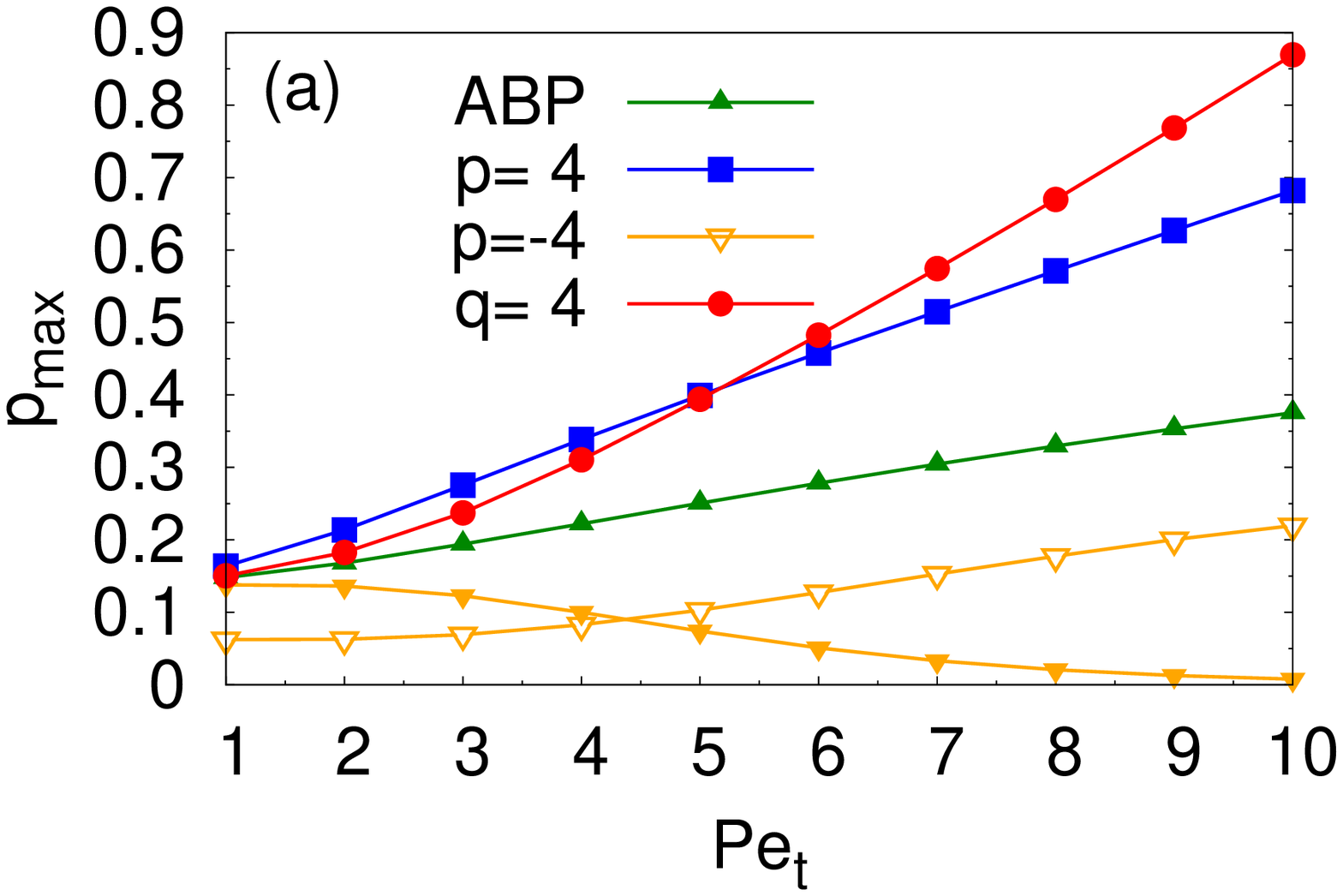}
 \includegraphics[scale=0.35]{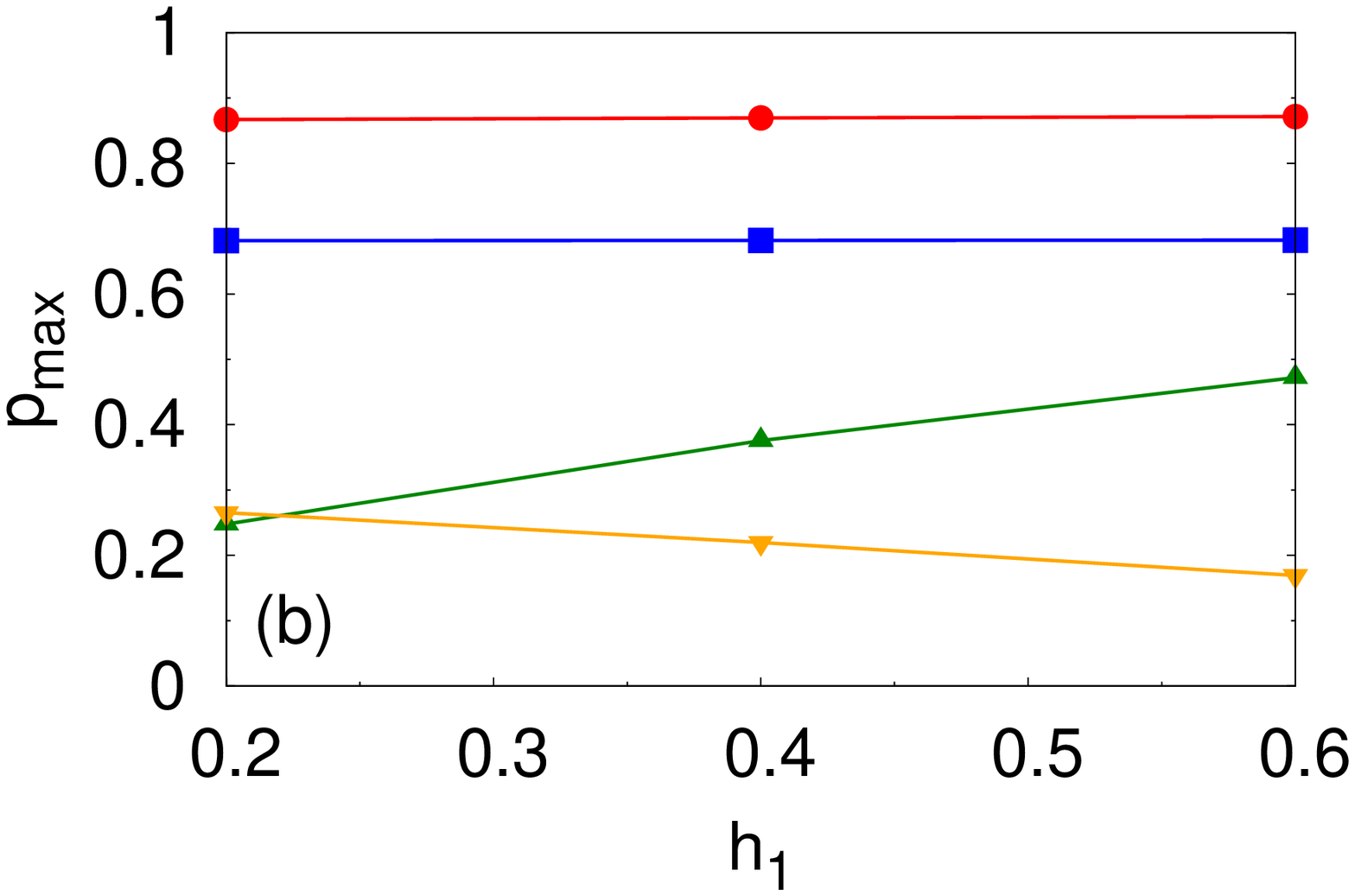}
\caption{Maximum value of the probability distribution, $p_{max}$, as a function of P\'eclet number (panel a,
$h_1 = 0.4$) and amplitude $h_1$ of the channel corrugation (panel b, $\mathrm{Pe}_t =10$).
For active Brownian particles (green upward triangles), neutral swimmers (red circles), and 
pushers (blue circles) the maximum occurs at $x=5$, whereas for pullers the maximum occurs at $x=0$, $x=5$ (orange open and filled downward triangles, respectively) depending on the value of $\mathrm{Pe}_t$. Pushers and pullers are characterized by $p=\pm 4$ and neutral swimmers by $q=4$.}
 \label{fig:max-symm}
\end{figure}
For $L_1=L/2$ the channel corrugation is symmetric hence the net current is zero, $I_x=0$. 
Figure~\ref{fig:symm_Pe} shows the behavior of $p(x)$, as obtained from Eq.~(\ref{eq:prob-corr-red}), as a function of the axial position $x$.
Interestingly, Fig.~\ref{fig:symm_Pe}(a) already shows a significant deviation in the axial density of an active Brownian particle compared to a passive one. All particles accumulate more at positions, where the channel is broadest, and the excess of accumulation increases strongly for larger values of $\mathrm{Pe}_t$. 
Looking at panels (b), (c) and (d) of Fig.~\ref{fig:symm_Pe}, we recognize that hydrodynamic coupling strongly modulates the distribution of swimmers as compared to active Brownian particles.
In particular,  neutral swimmers [panel (b)] and pushers [panel (c)] behave qualitatively similar to active Brownian particles, but the net effect of the hydrodynamic coupling induced by the walls is to enhance particle accumulation,
where the channel is broadest.
On the contrary, pullers [panel (d)] have quite a different behavior. Panel (d) of Fig.~\ref{fig:symm_Pe} shows that for smaller values of $\mathrm{Pe}_t$ pullers like the previous swimmers accumulate more in the broader parts of the channel. However, for increasing values of $\mathrm{Pe}_t$ the puller density becomes more uniform ($\mathrm{Pe}_t \simeq 3$) and, eventually, for $\mathrm{Pe}_t \geq 5$, very interestingly, pullers accumulate at channel bottlenecks. 

Figure\ \ref{fig:max-symm}(a) shows in detail the maximum of the probability distribution, $p_{max}$, as a function of 
$\mathrm{Pe}_t$. Clearly, for active Brownian particles, neutral swimmers, and pushers, the maximum $p_{max}$ increases monotonically upon increasing $\mathrm{Pe}_t$ and is always located at $x=L/2$. On the contrary, pullers are characterized by a  ``critical'' value of the P\'eclet number, $\mathrm{Pe}_c$, below which the maximum is located at $x=L/2$, whereas for $\mathrm{Pe}_t>\mathrm{Pe}_c$ the maximum is located at $x=0,L$. 
Interestingly, the value of $\mathrm{Pe}_c$ is properly captured by Eq.~(\ref{eq:Pe-C-3}), which indicates the P\'eclet number, where the swimmer dynamics crosses over from entropy-controlled to activity-controlled. 
It gives $\mathrm{Pe}_c\simeq 2$ for $h_1=0.4$ and $L=10$, which are the parameters used to obtain the data in Figs.~\ref{fig:symm_Pe} and \ref{fig:max-symm}.
Finally, we note that for pushers and neutral swimmers the value of $p_{max}$ is quite independent 
of the channel corrugation, $h_1$, whereas pullers and active Brownian particles are sensitive to $h_1$, as Fig.~\ref{fig:max-symm}(b) indicates.

When the channel is not symmetric, i.e., $L_1\neq L/2$ in Eq.~(\ref{eq:channel_shape}), in principle a non-zero net drift velocity $v_d$ [see Eq.~(\ref{eq:vel-x-corr})] occurs meaning rectification of the swimmer flow.
Therefore, the probability distribution $p(x)$ has to be calculated from the more general expression of Eq.~(\ref{eq:prob-corr}). 
In Fig.~\ref{fig:rhoVSzeta-asimm} we plot it for a range of $\mathrm{Pe}_t$ values for different swimmer types. The different graphs show that active Brownian particles as well as the other swimmer types behave similar as in a symmetric channel concerning the location of maximum accumulation and the variation with $\mathrm{Pe}_t$.
However, the probability distribution $p(x)$ reflects the asymmetry of the channel and gives a non-zero net drift. Indeed, as demonstrated in Fig.~\ref{fig:flux-asimm}, the different hydrodynamic swimmer types show a non-zero drift velocity $v_d$, while for active Brownian particles, which do not interact hydrodynamically with the channel walls, $v_d=0$. 
In particular, we find that both the sign and the magnitude of $v_d$ depends on the underlying physical mechanism leading to active displacement (hydrodynamic swimmer type) as well as on the value of $\mathrm{Pe}_t$ in a non-trivial way. 
Notably, we observe a non-monotonous behavior for all swimmers and for pullers also an inversion in the velocity. Finally the net swimmer flux vanishes if the channel becomes symmetric ($L_1=L/2$) as demonstrated in Fig.~\ref{fig:flux-asimm}(b).

\begin{figure}
\includegraphics[scale=0.3]{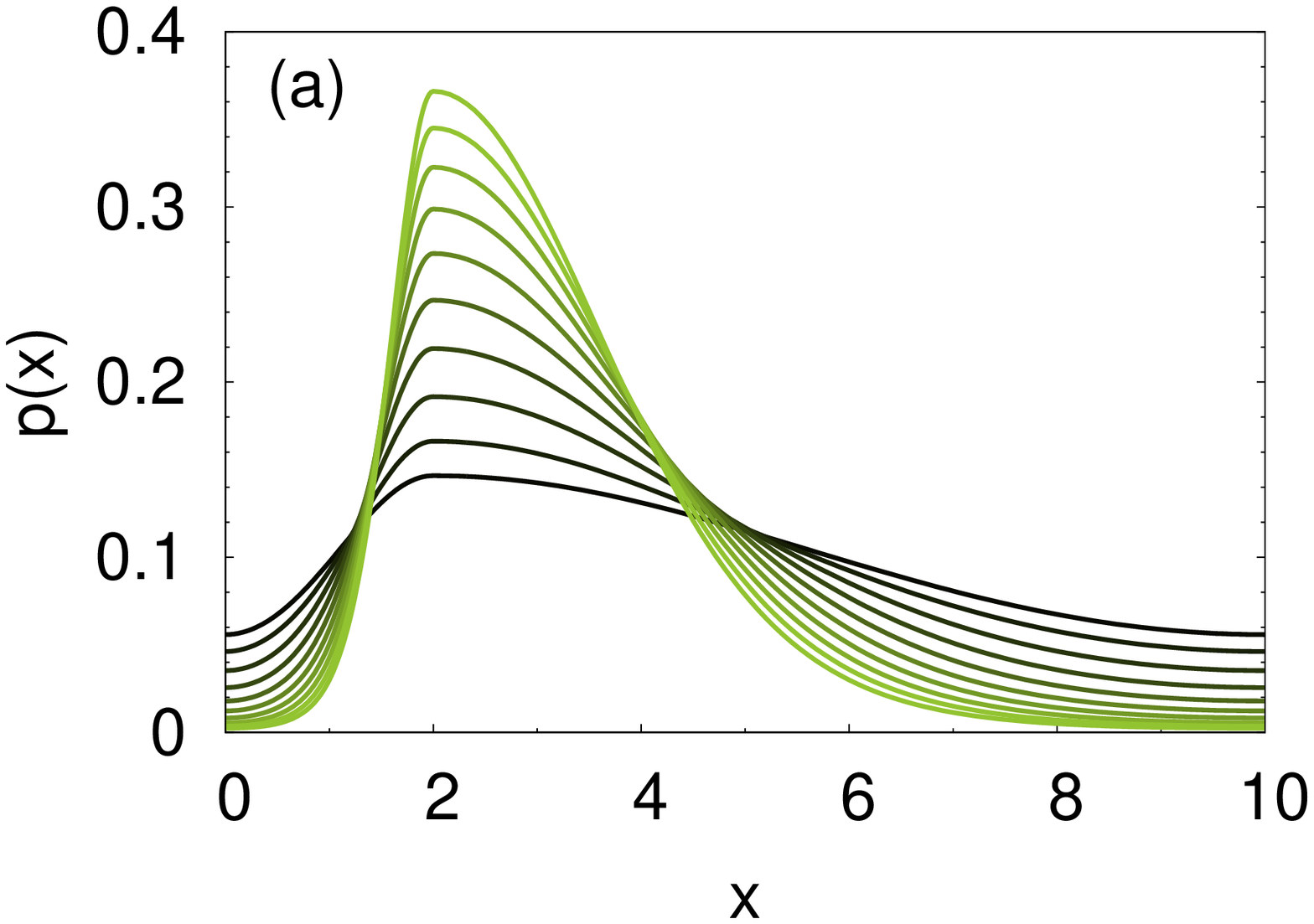}\, \includegraphics[scale=0.3]{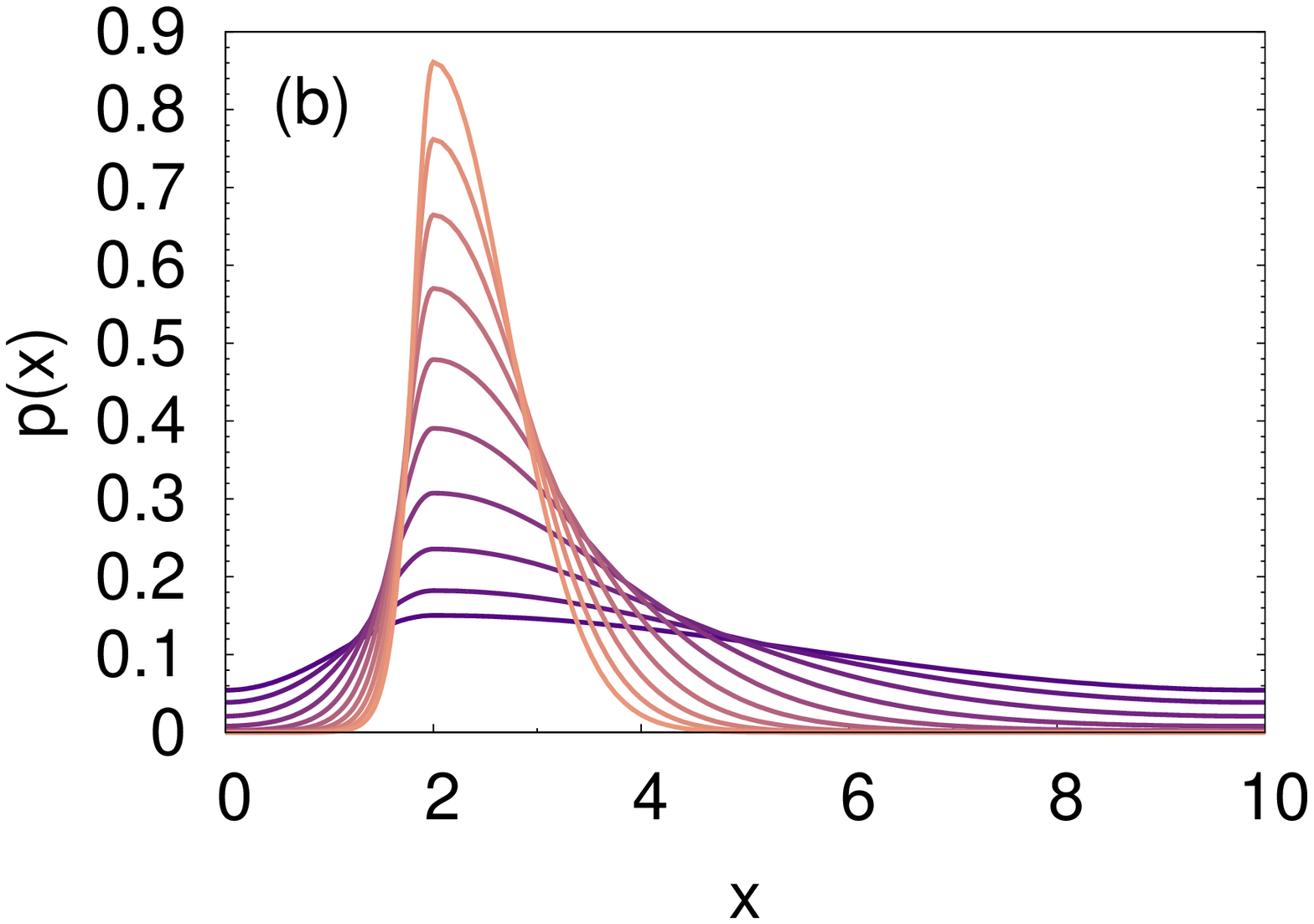}

\includegraphics[scale=0.3]{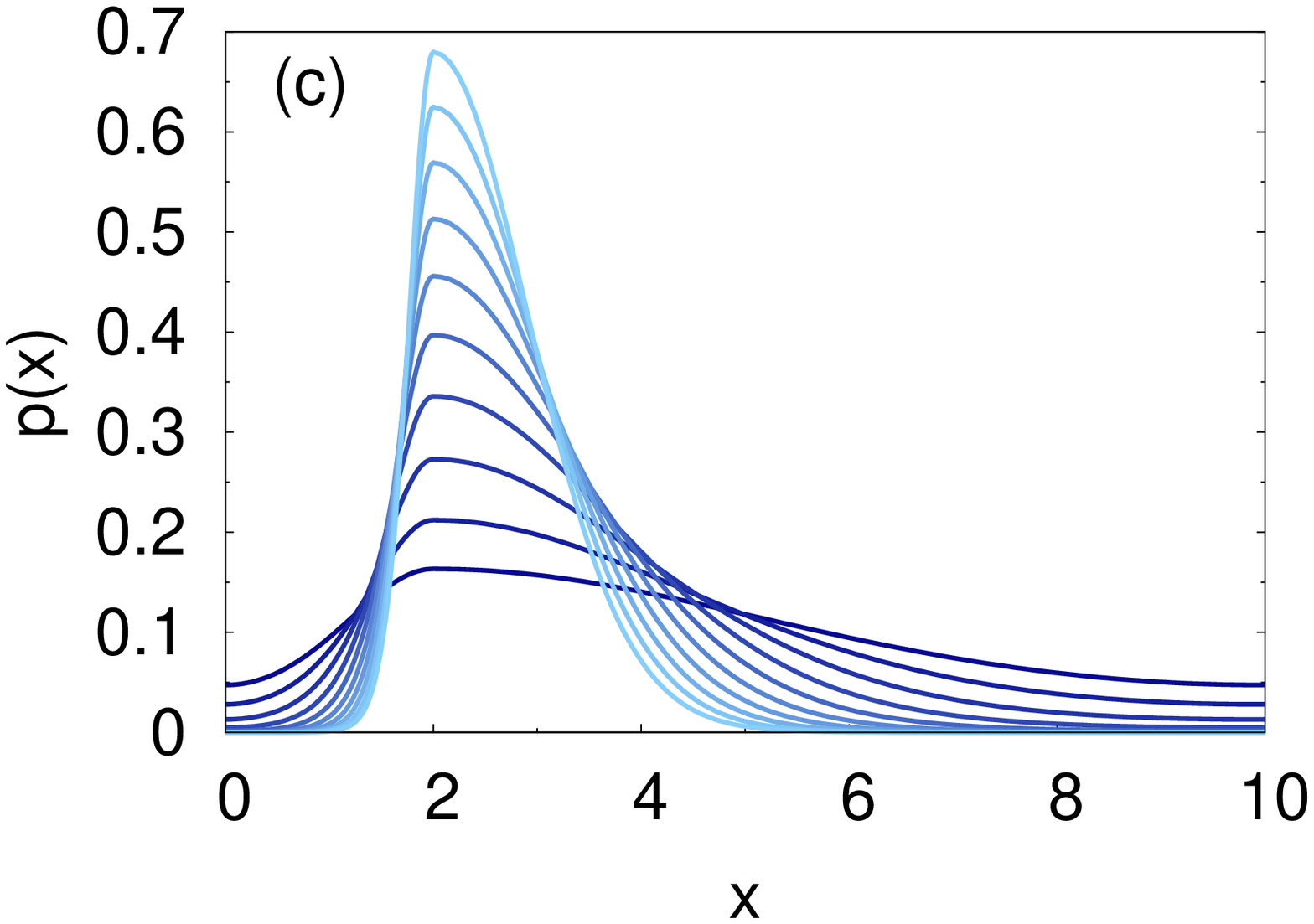}\, \includegraphics[scale=0.3]{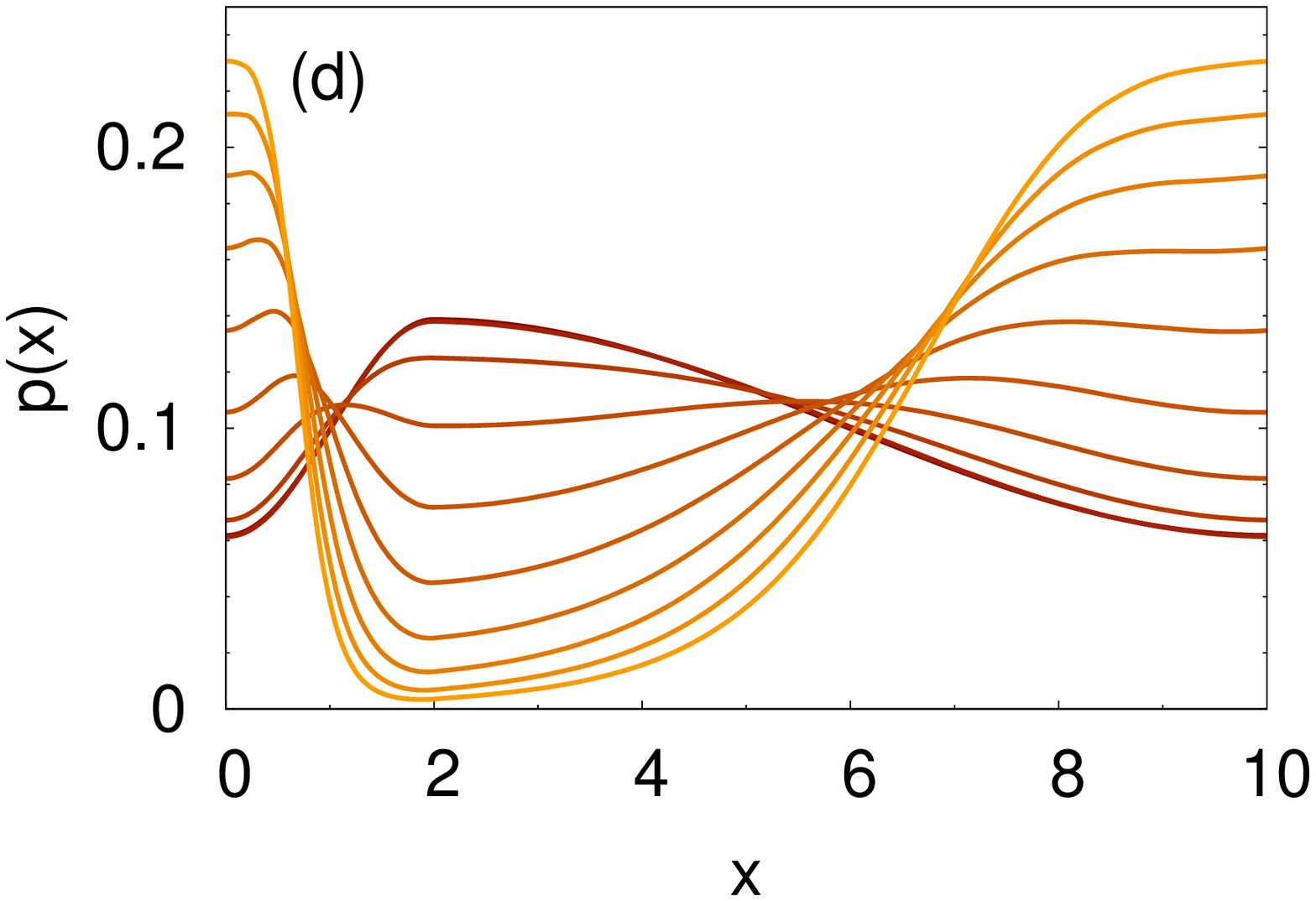}
\caption{Corrugated channel walls: asymmetric channel. Probability distribution function $p(x)$ of active Brownian particles (panel a), neutral swimmers ($q=4$, panel b), pushers ($p=4$, panel c), and pullers ($p=-4$, panel d) swimming in a channel characterized by $h_0=3$, $h_1=0.4$, $L=10$, and $L_1=0.2 L$ (all lengths in units of $d=h_0-R=R/2$). The different curves correspond to $\mathrm{Pe}_t=1,2,3,4,5,6,7,8,9$, and $10$, where the lines become fainter with increasing $\mathrm{Pe}_t$.}
\label{fig:rhoVSzeta-asimm}
\end{figure}

\begin{figure}
\includegraphics[scale=0.40]{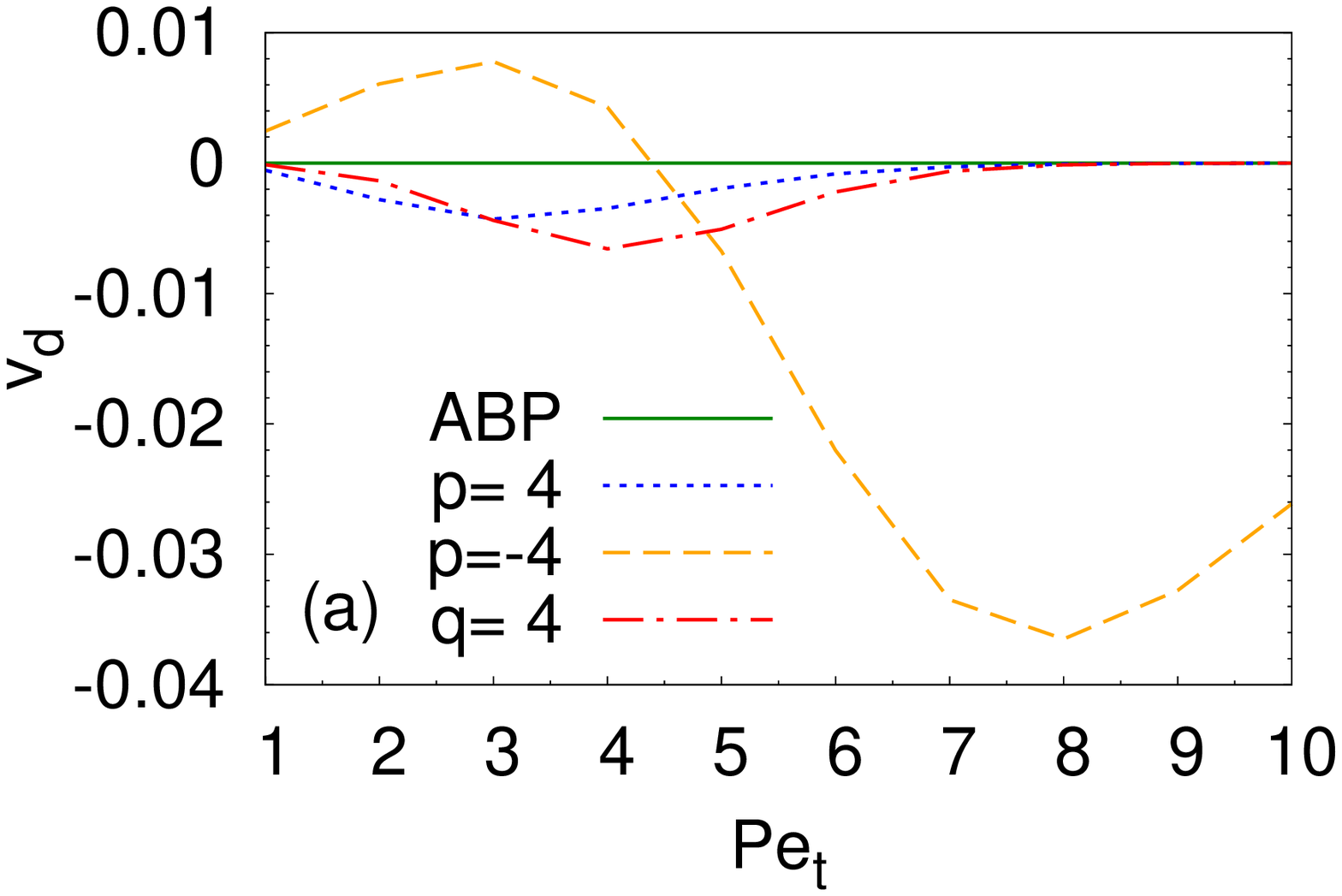}  \includegraphics[scale=0.40]{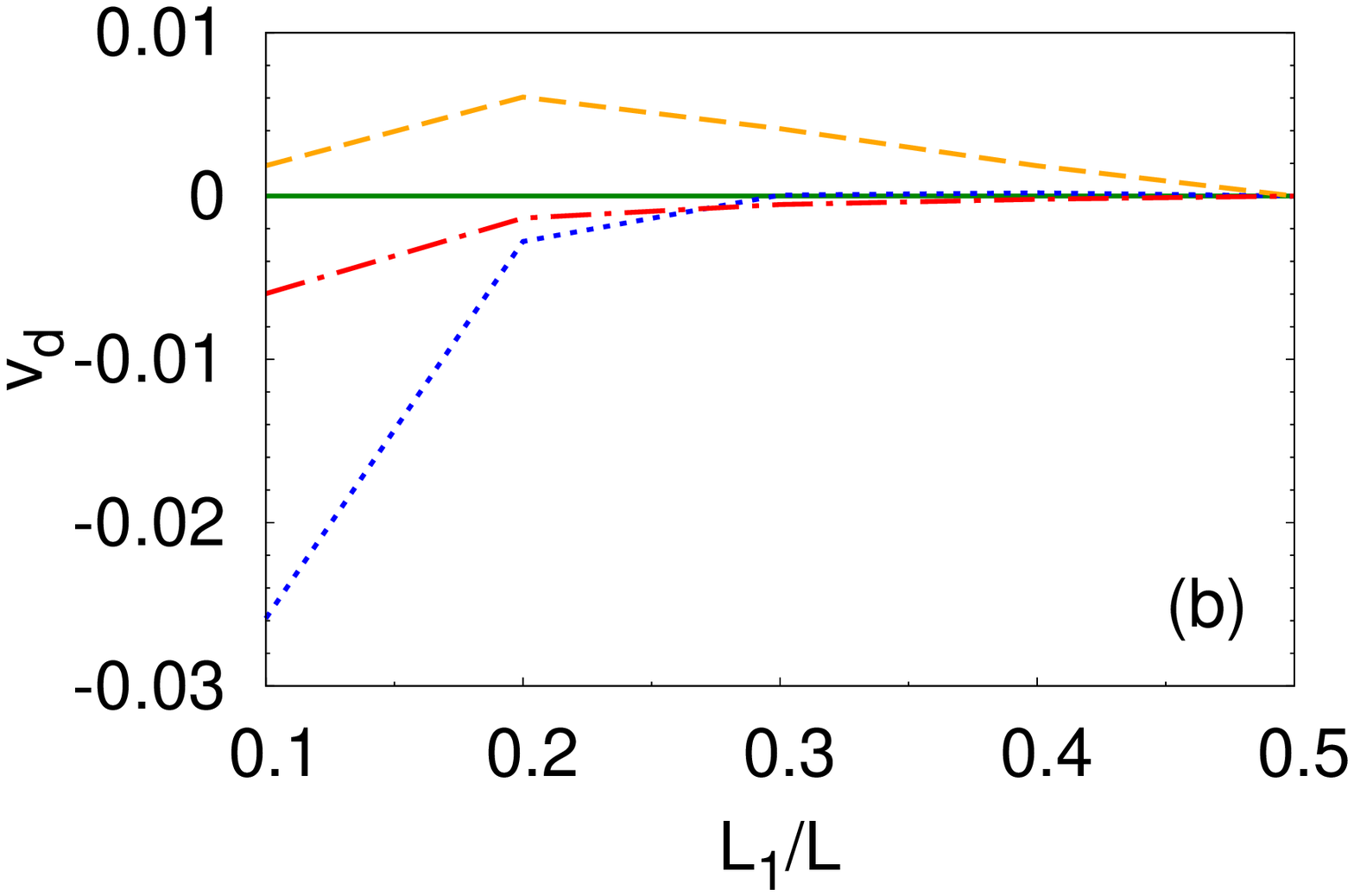}
 \caption{a) Net drift velocity $v_d$ as a function of the P\'eclet number $\mathrm{Pe}_t$ for pushers ($p=4$, blue dashed), pullers ($p=-4$, orange long-dashed), neutral ($q=4$, red dot-dashed), and active Brownian particles (green solid) in an asymmetric channel characterized by $h_1=0.4$ and $L_1=2$ (all lengths in units of $d=h_0-R=R/2$). b) Net drift velocity $v_d$ as a function $L_1$ for $\mathrm{Pe}_t=2$. Color code of the swimmer type is the same as in a).}
 \label{fig:flux-asimm}
\end{figure}

\section{Conclusions}
We have studied the dynamics of microswimmers in channels with corrugated walls. Under the assumption that the 
cross section of the channel varies smoothly, we have extended the Fick--Jacobs approximation to the case of active particles. 
Accordingly, the activity and the hydrodynamic coupling between the microswimmers and the channel walls can be captured by an effective potential 
$\chi(x)$ defined in Eq.~(\ref{eq:vel-x-corr}). Interestingly, our model identifies
two contributions to $\chi$, an active one proportional to $\mathrm{Pe}_t$ 
[first term on the right-hand side of Eq.~(\ref{eq:vel-x-corr})], and an entropic one [second term in 
Eq.~(\ref{eq:vel-x-corr})]. 
Unlike passive systems, the effective interaction between the microswimmers and the channel walls 
cannot be formulated with a conservative potential and, therefore, the entropic contribution to $\chi$ cannot be expressed in the form of Eq.~(\ref{eq:def-Ao}) familiar from passive systems.
In particular, our framework identifies a cross-over value, $\mathrm{Pe}_c$, of the P\'eclet number that separates two distinct regimes. 
For $\mathrm{Pe}_t\ll \mathrm{Pe}_c$ the system is in an entropy--controlled regime, in which the probability distribution is mainly determined by the geometry of the channel and thereby resembles the behavior  of passive particles. On the contrary, for $\mathrm{Pe}_t\gg \mathrm{Pe}_c$ the system switches to an activity--controlled regime, in which the probability distribution depends strongly on the details of the swimming mechanism. 

In order to quantify our findings, we have applied our framework to different types of \pa{model} microswimmers by taking into account their hydrodynamic interactions with the channel walls using the far--field approximation of Refs.~\cite{BrennerBook,Spagnolie2012}.
\pa{However, we also disregarded that the finite size of microswimmers disturbs these generic velocity fields.}
Under these assumptions, we have been able to characterize the behavior of \pa{our model} microswimmers in microchannels and to contrast it against active Brownian particles, i.e., active particles, which do not interact hydrodynamically with the walls.
Our results show that the effective dynamics of confined 
microswimmers is strongly affected by their swimming mechanism.
In fact, while pullers accumulate stronger than active Brownian particles at channel walls, 
for pushers and neutral swimmers accumulation is weaker. 
Such a behavior depending on the swimmer type can already be observed in channels with plane walls (see Fig.~\ref{fig:sm-rho-z}). 

For corrugated channels the behavior of different \pa{generic} microswimmer types is further amplified. Microswimmers show a non-uniform accumulation at the channel walls.
Active Brownian particles, pushers, and neutral swimmers accumulate preferentially at positions, where the channel cross section is widest, irrespective of the value of $\mathrm{Pe}_t$.
The accumulation is strongest for neutral swimmers followed by pushers and active Brownian particles. Pullers have a more involved dependence on $\mathrm{Pe}_t$. 
While for small values of $\mathrm{Pe}_t$ they also accumulate  in the regions of widest cross section, for $\mathrm{Pe}_t>\mathrm{Pe}_c$ they preferentially accumulate at channel bottlenecks (see Fig.~\ref{fig:symm_Pe}, Fig.~\ref{fig:max-symm}), which is a clear signature of active motion.
Finally, we have investigated the case, in which  the channel corrugation breaks the fore-aft symmetry. 
This initiates a net flux of 
microswimmers along the channel axis, the direction of which strongly depends on both the underlying swimming mechanism and the value of $\mathrm{Pe}_t$ that controls the hydrodynamic interactions with the walls.
 
Our work clearly demonstrates how the hydrodynamics of different swimmer types determines their dynamics in
channels with plane and corrugated channel walls. This has a clear potential for separating passive from active particles or different
microswimmers from each other, which, for example, is important in biomedical applications.

\section*{Acknowledgments}
P.M. acknowledges I. Pagonabarraga and P. Nowakowski for useful discussions and the Research Training Group GRK1558 funded by the Deutsche Forschungsgemeinschaft for travel grant and stay during scientific visiting at Technische Universit\"at Berlin.

\appendix

\section{Derivation of Eq.~(\ref{eq:def-A}) of the main text}\label{app:A}

Here we derive the last term in Eq.~(\ref{eq:def-A}). 
We divide the integral over $z$ into two parts:  
\begin{equation}\label{eq:deriv-h}
 \beta \int_{-\infty}^{\infty}\int_{-\pi}^{\pi} 
  \left[ \frac{\partial}{\partial x} C(x,z) \right]   
g(x,z,\theta) dzd\theta=\beta \int_{-\infty}^{0}\int_{-\pi}^{\pi} 
  \left[ \frac{\partial}{\partial x} C(x,z) \right]   
g(x,z,\theta) dzd\theta+\beta \int_{0}^{\infty}\int_{-\pi}^{\pi} 
  \left[ \frac{\partial}{\partial x} C(x,z) \right]   
g(x,z,\theta) dzd\theta
\end{equation}
In the following we 
show how to rewrite
the second term in the last expression. The first one can be 
treated
following the same approach.
In order to  compute the derivative of the potential $C(x,z)$, it is useful to 
introduce
the generalized Heaviside 
step
function $T(x)$:
\begin{equation}
 T(x)=\begin{cases}
        0 & z<h(x)-R-a\\
        \frac{1}{b+a}(z-h(x)+R+a) & h(x)-R-a\leq z \leq h(x)-R+b\\
        1 & z>h(x)-R+b
      \end{cases}
\end{equation}
where $a,b>0$ are small numbers\footnote{The choice of two different parameters, namely $a$ and $b$ for the regularization allows us to show that, as should be expected, the results do not depend on the value of the regularized distribution at $z=h(x)-R$.}, i.e. definitely smaller than $h(x)$. We remark that in the limit $a,b\rightarrow 0$ $T(x)$ recovers the Heaviside step function.
Accordingly, we rewrite $C(x,y)$ in its regularized form:
\begin{equation}\label{eq:reg-C}
 C(x,z)= \begin{cases}
         0 & z<h(x)-R-a\\
         \frac{\bar c}{b+a}(z-h(x)+R+a) & h(x)-R-a\leq z \leq h(x)-R+b\\
         \bar{c} & z>h(x)-R+b
        \end{cases}
\end{equation}
From the definition of the potential $W(x,z,\theta)$
in  
Eqs.~(\ref{eq:smol_sol_1}),(\ref{eq:def-W}), we have:
\begin{equation}\label{eq:reg-W}
 \mathrm{Pe}_tW(x,z,\pa{\theta})=\begin{cases}
         \mathrm{Pe}_t W_0(x,z,\theta) & z<h(x)-R-a\\
         \beta\frac{\bar c}{b+a}(z-h(x)+R+a)+\mathrm{Pe}_t W_0(x,z,\theta) & h(x)-R-a\leq z \leq h(x)-R+b\\
         \beta\bar{c}+\mathrm{Pe}_tW_0(x,h(x)-R+b,\theta) & z>h(x)-R+b
        \end{cases}
\end{equation}
where $W_0(x,z,\theta)=-z\cos\theta-\int v_{z}(z,\theta)dz$. We remark that in the limit $\bar{c}\rightarrow \infty$ and $a,b\rightarrow 0$, Eqs.~(\ref{eq:reg-C}),(\ref{eq:reg-W}) recover Eqs.~(\ref{eq:def_C}),(\ref{eq:def-W}).
According to Eq.~(\ref{eq:reg-C}) we have:
\begin{equation}
 \frac{\partial}{\partial x}C(x,z)=
 \begin{cases}
 0 &          z<h(x)-R-a\\                     
 -\bar{c}\frac{1}{b+a}\frac{\partial}{\partial x}h(x) & h(x)-R-a\leq z \leq h(x)-R+b\\
 0   & z>h(x)-R+b
 \end{cases}
\end{equation}
Thus,
the last integral in Eq.~(\ref{eq:deriv-h}) reads:
\begin{equation}
 \beta \int_{0}^{\infty}\int_{-\pi}^{\pi} 
  \left[ \frac{\partial}{\partial x} C(x,z) \right]   
g(x,z,\theta) dzd\theta=-\beta\bar{c}\frac{1}{b+a}\left[\frac{\partial}{\partial x}h(x)\right] \int_{h(x)-R-a}^{h(x)-R+b}\int_{-\pi}^{\pi} g(x,z,\theta) dzd\theta
\label{eq:app-int}
\end{equation}
Before calculating the integral in the r.h.s of Eq.~(\ref{eq:app-int}) we remark that in the limit $\bar{c}\rightarrow\infty$ the function $g(x,z,\theta)$ depends exponentially on $\bar{c}$ and therefore it is not possible to approximate the integral by the value of $g(x,z,\theta)$ at $z=h(x)-R$. Therefore, in order to calculate the integral in the r.h.s of Eq.~(\ref{eq:app-int}) we isolate the contributions in $g(x,z,\theta)$ that depend on $\bar{c}$ and integrate them separately from those contribution that are not dependent on $\bar c$.
Using Eq.(\ref{eq:g-corrugated}) we rewrite $g(x,z,\theta)$ as
\begin{equation}
g(x,z,\theta)=e^{-\mathrm{Pe}_tW(x,z,\theta)}\tilde g(x,\theta)
\end{equation}
Accordingly, using Eq.~(\ref{eq:reg-W}) the integral in Eq.~(\ref{eq:app-int}) gives:
\begin{equation}
 \int_{h(x)-R-a}^{h(x)-R+b}\int_{-\pi}^{\pi} g(x,z,\theta) dzd\theta=
 \int_{-\pi}^{\pi} d\theta \left[\tilde g(x,\theta)\int_{h(x)-R-a}^{h(x)-R+b}e^{-\beta\frac{\bar c}{b+a}(z-h(x)+R+a)-\mathrm{Pe}_tW_0(x,z,\theta)}dz\right]
\end{equation}
For small values of $a$ and $b$ we have $e^{-\mathrm{Pe}_tW_0(x,z,\theta)}\simeq e^{-\mathrm{Pe}_tW_0(x,h(x)-R,\theta)}$. Hence, taking $e^{-\mathrm{Pe}_tW_0(x,h(x)-R,\theta)}$ out of the integral in $z$ and performing the integration the last expression can be approximated by
\begin{eqnarray}
  \int_{-\pi}^{\pi}d\theta\left[ \tilde g(x,\theta) \int_{h(x)-R-a}^{h(x)-R+b}e^{-\pa{\beta}\frac{\bar c}{b+a}(z-h(x)+R+a)-\mathrm{Pe}_tW_0(x,z,\theta)}dz\right]&\simeq& \int_{-\pi}^{\pi}d\theta \Biggl[\tilde g(x,\theta)  e^{-\mathrm{Pe}_tW_0(x,h(x)-R,\theta)}\cdot\Biggr.\nonumber \\
  & &\cdot \left. \int_{h(x)-R-a}^{h(x)-R+b}e^{-\beta\frac{\bar c}{b+a}(z-h(x)+R+a)}dz\right]
  \label{eq:app-int1}
\end{eqnarray}
where the last integral leads to:
\begin{equation}
 \int_{h(x)-R-a}^{h(x)-R+b}e^{-\beta\frac{\bar c}{b+a}(z-h(x)+R+a)}dz=\frac{b+a}{\beta \bar c}\left(1-e^{-\beta \bar c}\right)
 \label{eq:app-int2}
\end{equation}
Finally substituting Eqs.~(\ref{eq:app-int1}),(\ref{eq:app-int2}) in Eq.~(\ref{eq:deriv-h}) we obtain:
\begin{equation}
 \beta \int_{-\infty}^{\infty}\int_{-\pi}^{\pi} 
  \left[ \frac{\partial}{\partial x} C(x,z) \right]   
g(x,z,\theta) dzd\theta=-2\left[\frac{\partial}{\partial x}h(x)\right] \left(1-e^{-\beta\bar c}\right)\int_{-\pi}^{\pi} g(x,h(x)-R,\theta) d\theta
\end{equation}
where we have used $g(x,h(x)-R,\theta)=\tilde g(x,\theta)e^{-\mathrm{Pe}_tW_0(x,h(x)-R,\pa{\theta})}$ and we have multiplied by $2$ since in Eq.~(\ref{eq:deriv-h}) there are two contributions on the rhs that, due to the axial symmetry of the channel provide an equal contribution to Eq.~(\ref{eq:deriv-h}).
Finally, in the limit $\bar c\rightarrow \infty$ we obtain:
\begin{equation}
 \beta \int_{-\infty}^{\infty}\int_{-\pi}^{\pi} 
  \left[ \frac{\partial}{\partial x} C(x,z) \right]   
g(x,z,\theta) dzd\theta=-2\left[\frac{\partial}{\partial x} h(x)\right]\int_{-\pi}^{\pi}  g(x,h(x)-R,\theta) d\theta
\end{equation}

\section{Derivation of the linear and angular velocity for channels with varying cross section}
\label{app:3}
In order to account for the constraints  imposed on the velocity field by the channel walls we exploit the method of images~\cite{BrennerBook,Spagnolie2012}. In the presence of two walls in principle we should consider an infinite series of images since the 
we have to take into account that the images introduced to fulfill the no-slip boundary conditions on a wall will affect the value of the 
velocity  field on the opposite wall. 
However, for the case under study at leading order in the far--field expansion the velocity field induced by microswimmers  decays as 
$1/r^2$ for pushers/pullers and $1/r^3$ for neutral swimmers, where $r$ is the distance from the particle. 
In order to derive analytical expressions we disregard the contributions of higher-order images and we approximate the series with its 
first-order contribution.
\begin{figure}[h]
\includegraphics[scale=0.5]{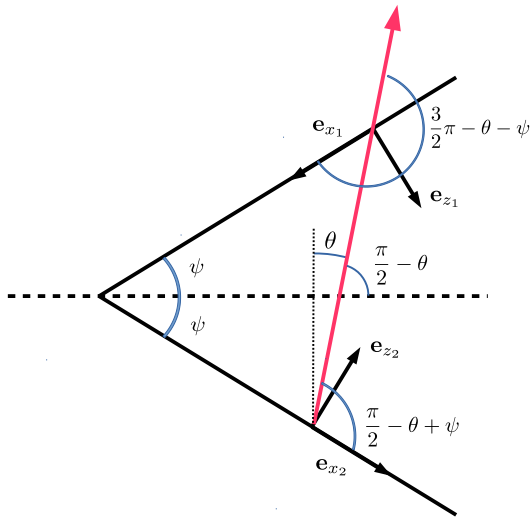} 
\caption{Reference frame. The red arrow represents the axis of symmetry of the swimmer.}
\label{fig:ref-frame}
\end{figure}
Under these assumptions, in the case of a channel whose cross section
varies on length scales much larger than particle size, we can locally approximate channel walls as flat planes tilted by an angle $\psi(x)=\arctan(\partial_x h(x))$ with respect to the channel longitudinal axis. 
According to Ref.~\cite{Spagnolie2012}, in dimensionless units, we have:
\begin{equation}
\omega(x,z,\theta)=\frac{R}{d}\left[-\frac{3p\sin(2\theta+2\psi(x))}{16\left(h_{0}-z\right)^{3}}+\frac{3q\sin(\theta+\psi(x))}{8\left(h_{0}-z\right)^{4}}+\frac{3p\sin(2\theta-2\psi(x))}{16\left(h_{0}+z\right)^{3}}-\frac{3q\sin(\theta-\psi(x))}{8\left(h_{0}+z\right)^{4}}\right]
\label{eq:omega-sqrm-1-1}
\end{equation}
where $p=p'/d^2$ and $q=q'/d^3$ are the dimensionless amplitudes of the force dipole and source dipole respectively.
When the channel cross section is not constant, the normal at the channel walls are not aligned with the $z$ direction. Therefore, the linear velocities read~\cite{Spagnolie2012}:
\begin{eqnarray*}
v_z(x,z,\theta) & = & v_{z,1}\mathbf{e}_{z,1}\cdot\mathbf{e}_{z}+v_{z,2}\mathbf{e}_{z,2}\cdot\mathbf{e}_{z}+v_{x,1}\mathbf{e}_{x,1}\cdot\mathbf{e}_{z}+v_{x,2}\mathbf{e}_{x,2}\cdot\mathbf{e}_{z}\\
v_x(x,z,\theta) & = & v_{z,1}\mathbf{e}_{z,1}\cdot\mathbf{e}_{x}+v_{z,2}\mathbf{e}_{z,2}\cdot\mathbf{e}_{x}+v_{x,1}\mathbf{e}_{x,1}\cdot\mathbf{e}_{x}+v_{x,2}\mathbf{e}_{x,2}\cdot\mathbf{e}_{x} \, ,
\end{eqnarray*}
where $v_{z,1,2}$ and $v_{x,1,2}$ are the corrections to the transverse
and longitudinal velocities due to the two walls.
Furthermore, 
$\mathbf{e}_{z,1,2}$ and $\mathbf{e}_{x,1,2}$ are the unit vectors along the $z$ and $x$
axis in the frame of reference 
aligned along the channel walls
(see Fig.~\ref{fig:ref-frame}), while $\mathbf{e}_{x,z}$ are the respective unit vectors along and perpendicular to the channel walls. In particular, we recall that 
\begin{eqnarray*}
& \mathbf{e}_{z,1}\cdot\mathbf{e}_{z} = -\cos(\psi(x)) & \mathbf{e}_{z,1}\cdot\mathbf{e}_{x}  =  \sin(\psi(x))\\
& \mathbf{e}_{z,2}\cdot\mathbf{e}_{z} =  \cos(\psi(x)) & \mathbf{e}_{z,2}\cdot\mathbf{e}_{x}  =  \sin(\psi(x))\\
& \mathbf{e}_{x,1}\cdot\mathbf{e}_{x} =  -\cos(\psi(x)) & \mathbf{e}_{x,1}\cdot\mathbf{e}_{z} =  -\sin(\psi(x))\\
& \mathbf{e}_{x,2}\cdot\mathbf{e}_{x} =  \cos(\psi(x))  & \mathbf{e}_{x,2}\cdot\mathbf{e}_{z} =  -\sin(\psi(x))
\end{eqnarray*}
Therefore we have:
\begin{eqnarray*}
v_z(x,z,\theta) & = & \cos(\psi(x))\left(v_{z,2}-v_{z,1}\right)-\sin(\psi(x))\left(v_{x,1}+v_{x,2}\right)\\
v_x(x,z,\theta) & = & \sin(\psi(x))\left(v_{z,1}+v_{z,2}\right)+\cos(\psi(x))\left(v_{x,2}-v_{x,1}\right)
\end{eqnarray*}
The magnitudes of the contributions are:
\begin{eqnarray*}
 v_{x,2}&=&\frac{3}{8}\frac{p}{\left(h_0+z\right)^2}\sin(2\theta-2\psi(x))-\frac{q}{4}\frac{1}{\left(h_0+z\right)^3}\sin(\theta-\psi(x))\\
 v_{x,1}&=&\frac{3}{8}\frac{p}{\left(h_0-z\right)^2}\sin(2\theta+2\psi(x))+\frac{q}{4}\frac{1}{\left(h_0-z\right)^3}\sin(\theta+\psi(x))\\
 v_{z,2}&=&-\frac{3}{8}\frac{p}{\left(h_0+z\right)^2}\left(1-3\cos^2(\theta-\psi(x))\right)-\frac{q}{\left(h_0+z\right)^3}\cos(\theta-\psi(x))\\
 v_{z,1}&=&-\frac{3}{8}\frac{p}{\left(h_0-z\right)^2}\left(1-3\cos^2(\theta+\psi(x))\right)+\frac{q}{\left(h_0-z\right)^3}\cos(\theta+\psi(x))\\
\end{eqnarray*}
Therefore the linear velocities read:
\begin{equation}
\begin{array}{c}
v_z(x,z,\theta)=\cos(\psi(x))\left[-\frac{3p}{8\left(h_{0}+z\right)^{2}}\left(1-3\cos^{2}(\theta-\psi(x))\right)-\frac{q}{\left(h_{0}+z\right)^{3}}\cos(\theta-\psi(x))+\right.\\
\left.+\frac{3p}{8\left(h_{0}-z\right)^{2}}\left(1-3\cos^{2}(\theta+\psi(x))\right)-\frac{q}{\left(h_{0}-z\right)^{3}}\cos(\theta+\psi(x))\right]\\
-\sin(\psi(x))\left[\frac{3p}{8\left(h_{0}+z\right)^{2}}\sin(2\theta-2\psi(x))-\frac{q}{4\left(h_{0}+z\right)^{3}}\sin(\theta-\psi(x))+\right.\\
\left.+\frac{3p}{8\left(h_{0}-z\right)^{2}}\sin(2\theta+2\psi(x))+\frac{q}{4\left(h_{0}-z\right)^{3}}\sin(\theta+\psi(x))\right]
\end{array}
\label{eq:val-sqrm-1-1}
\end{equation}
and 
\begin{equation}
\begin{array}{c}
v_x(x,z,\theta)=\cos(\psi(x))\left[\frac{3p}{8\left(h_{0}+z\right)^{2}}\sin(2\theta-2\psi(x))-\frac{q}{4\left(h_{0}+z\right)^{3}}\sin(\theta-\psi(x))+\right.\\
\left.-\frac{3p}{8\left(h_{0}-z\right)^{2}}\sin(2\theta+2\psi(x))-\frac{q}{4\left(h_{0}-z\right)^{3}}\sin(\theta+\psi(x))\right]+\\
\sin(\psi(x))\left[-\frac{3p}{8\left(h_{0}+z\right)^{2}}\left(1-3\cos^{2}(\theta-\psi(x))\right)-\frac{q}{\left(h_{0}+z\right)^{3}}\cos(\theta-\psi(x))+\right.\\
\left.-\frac{3p}{8\left(h_{0}-z\right)^{2}}\left(1-3\cos^{2}(\theta+\psi(x))\right)+\frac{q}{\left(h_{0}-z\right)^{3}}\cos(\theta+\psi(x))\right]
\end{array}
\label{eq:val-sqrm-1-1x}
\end{equation}
For plane channel walls we have $h(x)=h_0$ and $\psi(x)=0$. Accordingly Eqs.~(\ref{eq:omega-sqrm-1-1}),(\ref{eq:val-sqrm-1-1}),(\ref{eq:val-sqrm-1-1x}) 
reduce to:
\begin{eqnarray}
  \omega(x,z,\theta)&=&-3qR\sin(\theta)\frac{h_{0}z\left(h_{0}^{2}+z^{2}\right)}{\left(h_{0}^{2}-z^{2}\right)^{4}}-\frac{3}{8}pR\sin(2\theta)\frac{h_{0}\left(h_{0}^{2}+3z^{2}\right)}{\left(h_{0}^{2}-z^{2}\right)^{3}}\label{eq:omega-flat}\\
  v_z(x,z,\theta)&=&-2q\cos(\theta)\frac{h_0\left(h_{0}^{2}+3z^{2}\right)}{\left(h_{0}^{2}-z^{2}\right)^{3}}+\frac{3}{2}p\frac{h_{0}z}{\left(h_{0}^{2}-z^{2}\right)^{2}}\left(1-3\cos^{2}(\theta)\right)\label{eq:vz-flat}\\
  v_x(x,z,\theta)&=&-\frac{q}{2}\frac{h_0\left(h_0^2+3z^2\right)}{\left(h^2_{0}-z^2\right)^{3}}\sin(\theta)-\frac{3}{2}p\frac{h_0z}{\left(h^2_{0}-z^2\right)^{2}}\sin(2\theta)\label{eq:vx-flat}
\end{eqnarray}

\section{Enhanced hydrodynamic interactions for plane channel walls}\label{app:2}
\begin{figure}
\includegraphics[scale=0.45]{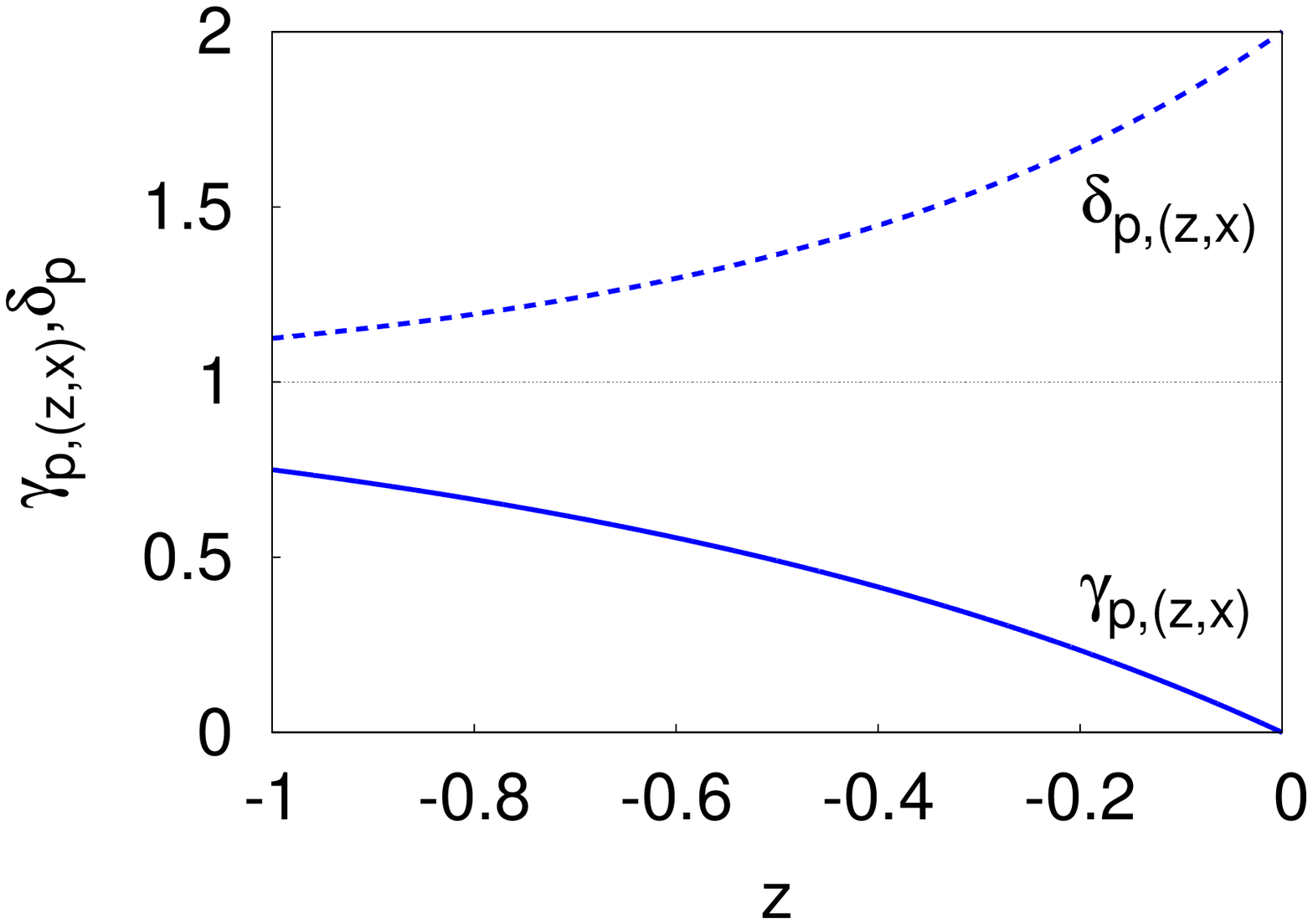} \includegraphics[scale=0.45]{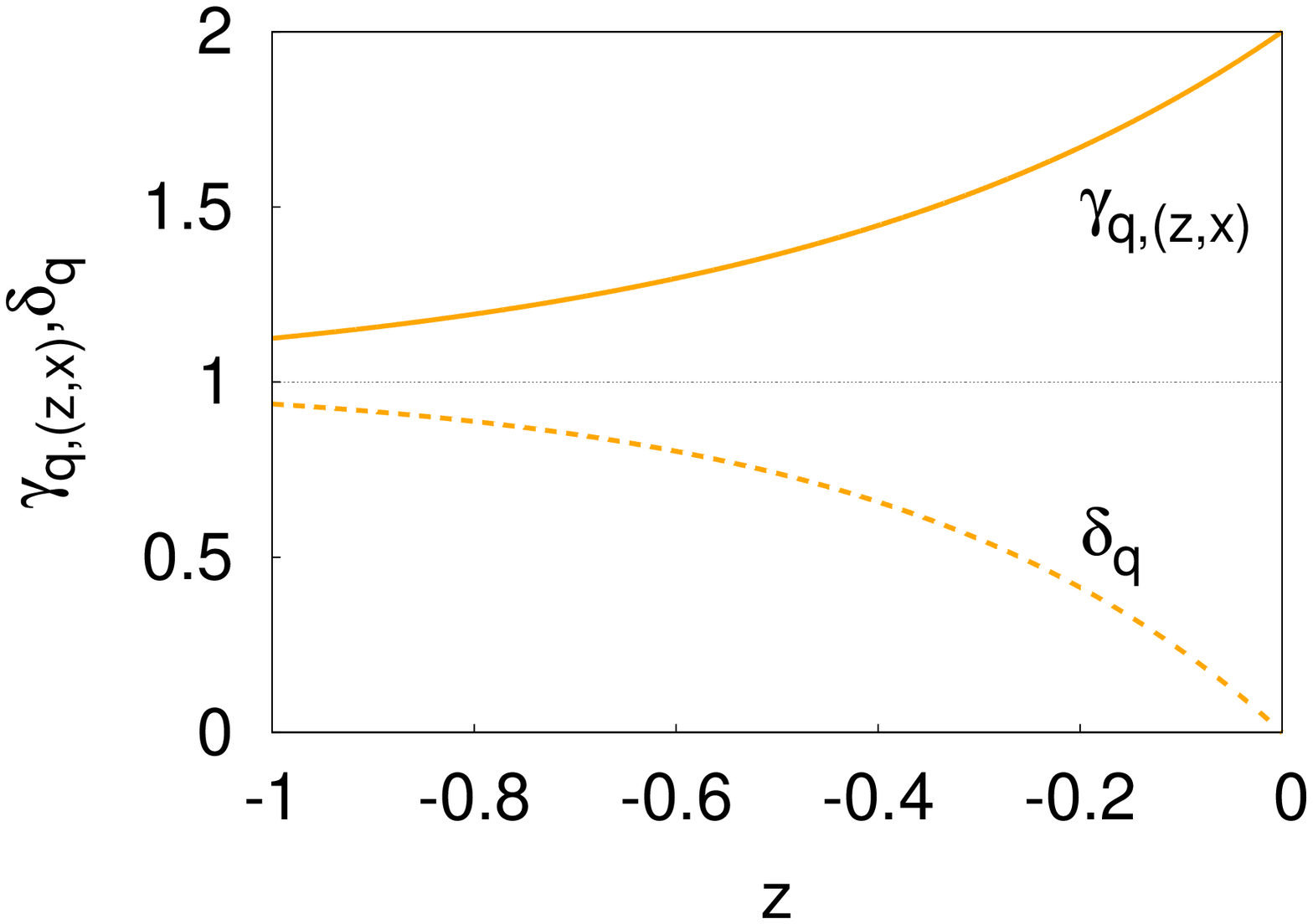}
\caption{
Linear and angular velocities for a swimmer confined between two parallel plates as a function of the distance from the
channel wall, normalized by that of a swimmer placed  at the same distance from a single planar wall. Solid (dashed) lines shows the ratio of linear (angular) velocities for a pusher/puller (left panel) and for a neutral swimmer (right panel). 
\label{fig:vel-ratio}}
\end{figure}
Eqs.~(\ref{eq:omega-flat})-(\ref{eq:vx-flat}) are derived by adding up the two contributions stemming from the hydrodynamic 
interactions with both walls. In order to discuss the relevance of these wall-induced velocities, we compare 
Eqs.~(\ref{eq:omega-flat})-(\ref{eq:vx-flat}) with the linear and angular velocity of a particle close to a single wall:
\begin{align} 
 \frac{\omega_{1W}R}{v_{0}}&=\frac{3q\sin(\theta)}{8\left(h_{0}+z\right)^{4}}\frac{R}{d}-\frac{3p\sin(2\theta)}{16\left(h_{0}+z\right)^{3}}\frac{R}{d}\\
 \frac{v_{z,1W}}{v_{0}}&=-\frac{q}{\left(h_{0}+z\right)^{3}}\cos(\theta)-\frac{3p}{8\left(h_{0}+z\right)^{2}}\left(1-3\cos^{2}(\theta)\right)\\
 \frac{v_{x,1W}}{v_0}&=-\frac{q}{4\left(h_{0}+z\right)^{3}}\sin(\theta)+\frac{3p}{8\left(h_{0}+z\right)^{2}}\sin(2\theta)
\end{align}
In particular, we are interested in the ratio between the contribution to the linear and angular velocity due to the hydrodynamic interactions with one or two walls, namely:
\begin{alignat}{3}
 &\delta_p=&&\frac{\omega(p,q=0)}{\omega_{1W}(p,q=0)}&&=2\frac{h_{0}\left(h_{0}^{2}+3z^{2}\right)}{\left(h_{0}-z\right)^{3}}\label{eq:delta-p}\\
 &\delta_q=&&\frac{\omega(p=0,q)}{\omega_{1W}(p=0,q)}&&=-8\frac{h_{0}z\left(h_{0}^{2}+z^{2}\right)}{\left(h_{0}-z\right)^{4}}
 \label{eq:delta-q}\\
 &\gamma_{p,(z,x)}=&&\frac{v_{(z,x)}(p,q=0)}{v_{(z,x),1W}(p,q=0)}&&=-4\frac{h_{0}z}{\left(h_{0}-z\right)^{2}} \\
 &\gamma_{q,(z,x)}=&&\frac{v_{(z,x)}(p=0,q)}{v_{(z,x),1W}(p=0,q)}&&=2\frac{h_{0}\left(h_0^2+3z^2\right)}{\left(h_{0}-z\right)^{3}}
\label{eq:gamma-q}
\end{alignat}
Interestingly, according to Eqs.~(\ref{eq:delta-p})-(\ref{eq:gamma-q}) the values of $\delta_{p,q}$ and $\gamma_{p,q}$ do not depend on the 
orientation of the microswimmers  for both pushers/pullers and neutral swimmers.
As shown in Fig.~\ref{fig:vel-ratio}, the two-walls scenario is quite different from the single-wall case. When particles are close to the wall (i.e. $z\rightarrow -1$ for the parameters of Fig.~\ref{fig:vel-ratio}), the contribution of the second wall vanishes and the value of $\gamma_{p,q}$ and $\delta_{p,q}$ approach unity. 
In contrast, when particles are closer to the center of the channel (i.e. $z\rightarrow 0$)  the presence of a second wall strongly affects the overall modulation in the linear and angular velocity as compared to the single wall case.

\bibliography{confined_swimmer}
\bibliographystyle{plainnat}
\end{document}